\documentclass[sigconf,screen]{acmart}

\newif\ifanonymous
\anonymousfalse
\relax

\AtBeginDocument{}

\usepackage[utf8]{inputenc} % allow utf-8 input
\usepackage[T1]{fontenc}    % use 8-bit T1 fonts
\usepackage{xcolor}         % colors
\definecolor{linkdarkblue}{rgb}{0, 0.08, 0.45}    % used for hyperlinks
\usepackage{url}            % simple URL typesetting
\usepackage{array}          % for table
\usepackage{colortbl}       % for table
\usepackage{makecell}       % for table
\usepackage{tabularx}       % for table
\usepackage{bigstrut}       % for bigstrut
\usepackage{multirow}       % for multirow
\usepackage{colortbl}       % for cell color
\usepackage{nicefrac}       % compact symbols for 1/2, etc.
\usepackage{microtype}      % microtypography
\usepackage{graphicx}       % for figures
\usepackage{bm}             % for bold math symbols
\usepackage[capitalise,noabbrev]{cleveref} % cleveref for \cref
\usepackage{subcaption}     % for subfigures
\usepackage{booktabs}       % for tables
\usepackage{wrapfig}        % for wrapfigure
\usepackage{pifont}         % for checkmark
\usepackage{algorithm}      % for algorithm
\usepackage{mathtools}      % for math
\usepackage{enumitem}       % for enumerate
\usepackage{pifont}         % for more symbols
\usepackage{algorithm}      % for algorithm
\usepackage{algorithmicx}   % for algorithm
\usepackage{algpseudocode}  % for algorithm
\usepackage{thm-restate}    % for restatable theorems
\usepackage{fontawesome}    % for icons
\usepackage{comment}        % for comment content
\usepackage[normalem]{ulem} % for underline
\usepackage{amsmath,amsfonts,bm}

\def\eqref#1{equation~\ref{#1}}
\def\1{\bm{1}}

\DeclareMathAlphabet{\mathsfit}{\encodingdefault}{\sfdefault}{m}{sl}
\SetMathAlphabet{\mathsfit}{bold}{\encodingdefault}{\sfdefault}{bx}{n}

\allowdisplaybreaks

\newcommand{\noindentparnolinenospace}[1]{\noindent\textbf{#1} }

\newcommand{\noindentparnoline}[1]{\addvspace{0.50\baselineskip}\noindent\textbf{#1} }

\newcommand{\eqnref}[1]{(\ref{#1})}

\numberwithin{equation}{section}    % number equations within sections
\numberwithin{algorithm}{section}   % number algorithms within sections

\theoremstyle{plain}

\newtheorem*{theorem*}{Theorem}

\newtheorem*{lemma*}{Lemma}

\newtheorem*{axiom*}{Axiom}

\newtheorem*{conjecture*}{Conjecture}

\newtheorem*{assumption*}{Assumption}

\newtheorem*{proposition*}{Proposition}

\newtheorem*{corollary*}{Corollary}

\theoremstyle{definition}

\newtheorem*{definition*}{Definition}

\newtheorem*{example*}{Example}

\newtheorem*{problem*}{Problem}

\newtheorem*{question*}{Question}

\newtheorem*{exercise*}{Exercise}

\newtheorem*{remark*}{Remark}

\definecolor{darkgreen}{rgb}{0.0, 0.5, 0.0}
\definecolor{skyblue}{rgb}{0.53, 0.81, 0.98}
\definecolor{beamsearchblue}{rgb}{0.42, 0.56, 0.75}     % (108, 142, 191)
\definecolor{beamsearchyellow}{rgb}{0.84, 0.71, 0.34}   % (214, 182, 86)
\definecolor{beamsearchred}{rgb}{0.72, 0.33, 0.31}      % (184, 84, 80)
\definecolor{darkred}{rgb}{0.6, 0, 0}
\definecolor{pruning_red}{rgb}{0.64, 0.12, 0.21}        % (163, 31, 53)
\definecolor{pruning_gray}{rgb}{0.6, 0.6, 0.6}          % (153, 153, 153)
\definecolor{sensitivity_blue}{rgb}{0.30, 0.45, 0.69}   % (77, 115, 176)
\definecolor{sensitivity_orange}{rgb}{0.87, 0.52, 0.32} % (222, 133, 82)

\newcolumntype{Y}{>{\centering\arraybackslash}X}

\newcommand{\ie}{\text{i.e., }}         % that is
\newcommand{\eg}{\text{e.g., }}         % for example
\newcommand{\cf}{\text{cf. }}           % compare with or consult
\newcommand{\aka}{\text{a.k.a. }}       % also known as
\newcommand{\versus}{\text{vs. }}       % versus
\copyrightyear{2026}
\acmYear{2026}
\setcopyright{cc}
\setcctype{by-nc-nd}
\acmConference[SIGIR '26]{Proceedings of the 49th International ACM SIGIR Conference on Research and Development in Information Retrieval}{July 20--24, 2026}{Melbourne, VIC, Australia}
\acmBooktitle{Proceedings of the 49th International ACM SIGIR Conference on Research and Development in Information Retrieval (SIGIR '26), July 20--24, 2026, Melbourne, VIC, Australia}
\acmDOI{10.48550/arXiv.2601.22925}
\acmISBN{979-8-4007-2599-9/2026/07}

\begin{document}

%%%%%%%%%%%%%%%%%%%%%%%%%%%%%%%%%%%%%%%%%%%%%%%%%%%%%%%%%%%%%%%%%%%%%
% TITLE -------------------------------------------------------------
% The "title" command has an optional parameter,
% allowing the author to define a "short title" to be used in page headers.
\title{
BEAR: Towards Beam-Search-Aware Optimization for Recommendation with Large Language Models
}

% The "author" command and its associated commands are used to define
% the authors and their affiliations.
% Of note is the shared affiliation of the first two authors, and the
% "authornote" and "authornotemark" commands
% used to denote shared contribution to the research.

\author{Weiqin Yang}
    \orcid{0000-0002-5750-5515}
    \affiliation{
        \institution{Zhejiang University}
        \city{Hangzhou}
        \country{China}
    }
    \email{tinysnow@zju.edu.cn}
\authornotemark[2]
\authornotemark[3]

\author{Bohao Wang}
    \orcid{0009-0006-8264-3182}
    \affiliation{
        \institution{Zhejiang University}
        \city{Hangzhou}
        \country{China}
    }
    \email{bohao.wang@zju.edu.cn}
\authornotemark[2]
\authornotemark[3]

\author{Zhenxiang Xu}
	\orcid{0009-0004-3678-1141}
    \affiliation{
        \institution{Zhejiang University}
        \city{Hangzhou}
        \country{China}
    }
	\email{zhenxiangxu@zju.edu.cn}
\authornotemark[2]
\authornotemark[3]

\author{Jiawei Chen}
    \orcid{0000-0002-4752-2629}
    \affiliation{
        \institution{Zhejiang University}
        \city{Hangzhou}
        \country{China}
    }
    \email{sleepyhunt@zju.edu.cn}
\authornote{Corresponding author.}
\authornote{State Key Laboratory of Blockchain and Data Security, Zhejiang University.}
\authornote{College of Computer Science and Technology, Zhejiang University.}
\authornote{Hangzhou High-Tech Zone (Binjiang) Institute of Blockchain and Data Security.}

\author{Shengjia Zhang}
    \orcid{0009-0004-0209-2276}
    \affiliation{
        \institution{Zhejiang University}
        \city{Hangzhou}
        \country{China}
    }
    \email{shengjia.zhang@zju.edu.cn}
\authornotemark[2]
\authornotemark[3]

\author{Jingbang Chen}
    \orcid{0000-0002-7279-0801}
    \affiliation{
        \institution{The Chinese University of Hong Kong, Shenzhen}
        \city{Shenzhen}
        \country{China}
    }
    \email{chenjb@cuhk.edu.cn}

\author{Canghong Jin}
    \orcid{0000-0002-9774-9688}
    \affiliation{
        \institution{Hangzhou City University}
        \city{Hangzhou}
        \country{China}
    }
    \email{jinch@zucc.edu.cn}

\author{Can Wang}
    \orcid{0000-0002-5890-4307}
    \affiliation{
        \institution{Zhejiang University}
        \city{Hangzhou}
        \country{China}
    }
    \email{wcan@zju.edu.cn}
\authornotemark[2]
\authornotemark[4]

% By default, the full list of authors will be used in the page
% headers. Often, this list is too long, and will overlap
% other information printed in the page headers. This command allows
% the author to define a more concise list
% of authors' names for this purpose.
\ifanonymous
    % do nothing
\else
    \renewcommand{\shortauthors}{Weiqin Yang et al.}
\fi

% END: TITLE -------------------------------------------------------
%%%%%%%%%%%%%%%%%%%%%%%%%%%%%%%%%%%%%%%%%%%%%%%%%%%%%%%%%%%%%%%%%%%%%

%%%%%%%%%%%%%%%%%%%%%%%%%%%%%%%%%%%%%%%%%%%%%%%%%%%%%%%%%%%%%%%%%%%%%
% ABSTRACT ----------------------------------------------------------

\begin{abstract}
    Recent years have seen a rapid surge in research leveraging Large Language Models (LLMs) for recommendation. These methods typically employ supervised fine-tuning (SFT) to adapt LLMs to recommendation scenarios, and utilize beam search during inference to efficiently retrieve $B$ top-ranked recommended items. However, we identify a critical \emph{training-inference inconsistency}: while SFT optimizes the overall probability of positive items, it does not guarantee that such items will be retrieved by beam search even if they possess high overall probabilities. Due to the greedy pruning mechanism, beam search can prematurely discard a positive item once its prefix probability is insufficient.

    To address this inconsistency, we propose \textbf{BEAR} (\textbf{\uline{B}}eam-S\textbf{\uline{E}}arch-\textbf{\uline{A}}ware \textbf{\uline{R}}egularization), a novel fine-tuning objective that explicitly accounts for beam search behavior during training. Rather than directly simulating beam search for each instance during training, which is computationally prohibitive, BEAR enforces a relaxed \emph{necessary condition}: each token in a positive item must rank within the top-$B$ candidate tokens at each decoding step. This objective effectively mitigates the risk of incorrect pruning while incurring negligible computational overhead compared to standard SFT. Extensive experiments across four real-world datasets demonstrate that BEAR significantly outperforms strong baselines. Code is available at \url{https://github.com/Tiny-Snow/BEAR-SIGIR-2026}.
\end{abstract}



% Keywords. The author(s) should pick words that accurately describe
% the work being presented. Separate the keywords with commas.

% A "teaser" image appears between the author and affiliation
% information and the body of the document, and typically spans the
% page.

% \received{Day Month Year}
% \received[revised]{Day Month Year}
% \received[accepted]{Day Month Year}

% This command processes the author and affiliation and title
% information and builds the first part of the formatted document.
\maketitle

% END: ABSTRACT -----------------------------------------------------
%%%%%%%%%%%%%%%%%%%%%%%%%%%%%%%%%%%%%%%%%%%%%%%%%%%%%%%%%%%%%%%%%%%%%

%%%%%%%%%%%%%%%%%%%%%%%%%%%%%%%%%%%%%%%%%%%%%%%%%%%%%%%%%%%%%%%%%%%%%
% INTRODUCTION ------------------------------------------------------

\section{Introduction} \label{sec:introduction}

Large Language Models (LLMs) have exhibited remarkable abilities in natural language understanding~\citep{brown2020language}, generation~\citep{radford2019language}, and reasoning~\citep{wei2022chain}, achieving notable successes across a wide range of domains~\citep{rajpurkar2016squad,zhu2024deepseek,yang2024qwen2,huang2026wese}. These advances have sparked growing interest in leveraging LLMs as the backbone for recommender systems, \aka LLM-based RS~\citep{hou2024large,geng2022recommendation,shi2024large,zhang2025recommendation,li2023exploring,cui2022m6,zhang2025reinforced,cui2025field,cui2025hatllm,cui2026spectran}. This paradigm typically reformulates recommendation as a natural language task, involving three key stages: (i) \emph{Prompt construction}, converting a user's historical interactions into a structured textual prompt to guide the LLMs; (ii) \emph{Supervised fine-tuning} (SFT), pairing prompts with target positive items as training instances to adapt the LLMs for recommendation; and (iii) \emph{Inference}, generating $B$ candidate items via decoding strategies, and selecting the top-$K$ ranked items as final recommendations~\citep{bao2024decoding,lin2025igd,freitag2017beam}. Compared with conventional recommenders, LLM-based RS can perform fine-grained preference modeling~\citep{liao2024rosepo}, understand complex contextual information~\citep{bao2023tallrec}, and enhance recommendation interpretability~\citep{gao2023chat}, making them a promising and emerging direction.

% FIGURE: INTRODUCTION FIG ------------------------------------------
\begin{figure*}[t]
    \centering
    \includegraphics[width=\textwidth]{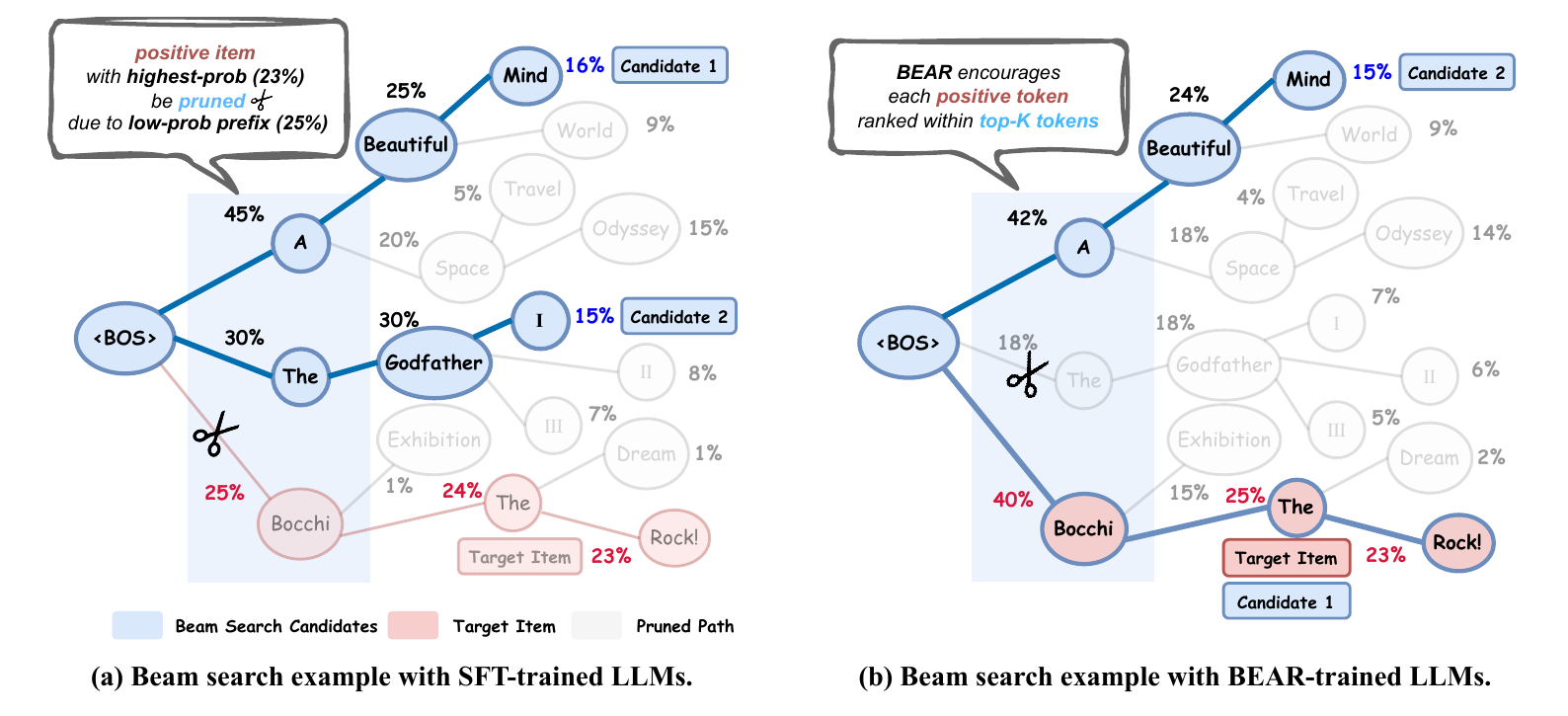}
    \caption{
        Illustration of beam search applied to SFT-trained LLMs \versus BEAR-trained LLMs (with beam width $B = 2$). The \textcolor{beamsearchblue}{blue paths} represent \textcolor{beamsearchblue}{candidate items} retrieved by beam search (\eg "\textcolor{beamsearchblue}{A Beautiful Mind}"), while \textcolor{beamsearchred}{red circles} mark the \textcolor{beamsearchred}{positive item} (\eg "\textcolor{beamsearchred}{Bocchi the Rock!}"). (a) In the SFT case, although the positive item possesses the highest overall probability, it is pruned at the first step due to the insufficient probability of its prefix (\eg "\textcolor{beamsearchred}{Bocchi}"). (b) BEAR promotes ranking each positive token within the top-$B$ among all possible tokens, making beam search more likely to retain the positive item.
    }
    \Description{Beam search examples with SFT and BEAR.}
    \label{fig:intro}
\end{figure*}
% END FIGURE: INTRODUCTION FIG --------------------------------------

During inference, beam search is widely adopted to efficiently retrieve items with high probabilities~\citep{bao2024decoding,lin2025igd,wang2025msl}. Beam search incrementally extends sequences token-by-token, retaining only a small number (\ie the beam width $B$) of the highest-probability candidates at each step, while discarding the others. This approach is considerably faster than exhaustive retrieval over all items, but it introduces a \emph{greedy pruning} mechanism: if a positive item's prefix has relatively low probability, it may be pruned early, even if its complete sequence has the highest overall item probability.

\begin{figure}[t]
    \centering
    \includegraphics[width=0.8\columnwidth]{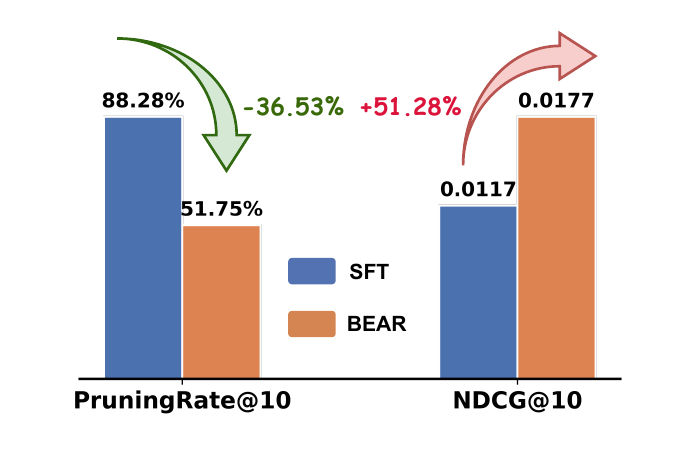}
    \caption{
        Performance comparison between SFT and BEAR on the Book dataset. PruningRate@10 denotes the pruning rate with beam width $B = 10$, \ie the proportion of positive items that rank within the top-10 in overall probability yet are pruned during beam search. The complete results are presented in \cref{tab:main-results,fig:pruning-rate}.
    }
    \Description{Pruning rate and NDCG comparison on Book dataset.}
    \label{fig:intro-ratio}
\end{figure}

\noindentparnoline{Challenges.}
We identify a fundamental \emph{training-inference inconsistency} between SFT optimization and beam search decoding in current LLM-based RS. While SFT targets improving the overall probability of positive items, this objective does not necessarily ensure that the item would be retrieved during beam search. \cref{fig:intro}(a) illustrates an example: despite the positive item "\textcolor{beamsearchred}{Bocchi the Rock!}" achieving the highest overall probability, its low-probability prefix "\textcolor{beamsearchred}{Bocchi}" causes premature pruning during beam search. Our empirical analysis on real-world datasets reveals that this phenomenon is alarmingly prevalent: over 80\% of positive items with top-$B$ overall probability are pruned before reaching the final recommendations of beam search (\cf \cref{fig:intro-ratio}). This striking statistic demonstrates that the training-inference inconsistency constitutes the primary bottleneck in LLM-based RS, demanding urgent resolution.

\noindentparnoline{Our Method.}
To bridge this gap, we propose a novel fine-tuning objective, termed \textbf{BEAR} (\textbf{\uline{B}}eam-S\textbf{\uline{E}}arch-\textbf{\uline{A}}ware \textbf{\uline{R}}egularization), which explicitly incorporates the beam search dynamics into training. A straightforward yet naive approach would be to simulate beam search for every instance during training, increasing the \emph{prefix} probabilities of positive items until they rank in the top-$B$ candidate prefixes at each step. However, this approach entails multiple forward passes per instance during training, making it computationally infeasible (\cf \cref{fig:runtime}). Instead, BEAR optimizes a relaxed \emph{necessary condition} for retrieving positive items via beam search, \ie ensuring each \emph{token} of the positive items ranks within the top-$B$ candidate tokens. Our motivation stems from an interesting empirical observation: violating this necessary condition constitutes the primary cause of incorrect pruning, accounting for over 70\% of cases in real-world datasets (\cf \cref{fig:pr_and_tpr_pruning_rate}). Therefore, optimizing BEAR would naturally and substantially mitigate the risk of incorrect pruning, thereby improving recommendation performance (\cf \cref{fig:intro-ratio}). Moreover, unlike the straightforward solution that simulates beam search, BEAR requires no additional forward passes, achieving high computational efficiency with negligible additional overhead compared to standard SFT, making it highly practical for real-world applications (\cf \cref{fig:runtime}).

To empirically validate the effectiveness of BEAR, we conduct comprehensive experiments on four real-world datasets.
Experimental results demonstrate that BEAR surpasses nine state-of-the-art fine-tuning baselines for LLM-based RS (mainly published within the past year), with \emph{a notable average improvement of 12.50\%} in recommendation accuracy across all datasets. Moreover, BEAR is a model-agnostic fine-tuning objective and can be seamlessly integrated into various LLM-based RS backbones. Empirically, we evaluate BEAR on three well-known recommendation backbones and obtain consistent performance gains. Additional analyses indicate that BEAR can indeed reduce the ratio of incorrect pruning, thereby validating its effectiveness in addressing the training-inference inconsistency in LLM-based RS.

\clearpage
\noindentparnoline{Contributions.}
In summary, our contributions are as follows:

\begin{itemize}[topsep=3pt,leftmargin=10pt,itemsep=0pt]
    \item We identify and analyze the critical issue of training-inference inconsistency in LLM-based recommendation tasks, and advocate explicitly incorporating the impact of beam search into the design of fine-tuning objectives.
    \item We propose BEAR, a novel beam-search-aware fine-tuning objective that optimizes a necessary condition for retrieving positive items via beam search. This objective effectively mitigates the risk of incorrect pruning in beam search with negligible additional computational overhead.
    \item We conduct extensive experiments to validate the effectiveness of BEAR, achieving a notable average improvement of 12.50\% over state-of-the-art methods across multiple datasets and backbones.
\end{itemize}

% END: INTRODUCTION -------------------------------------------------
%%%%%%%%%%%%%%%%%%%%%%%%%%%%%%%%%%%%%%%%%%%%%%%%%%%%%%%%%%%%%%%%%%%%%

% FIGURE: PROMPT ----------------------------------------------------
\begin{figure}[t]
    \centering
    \includegraphics[width=0.95\columnwidth]{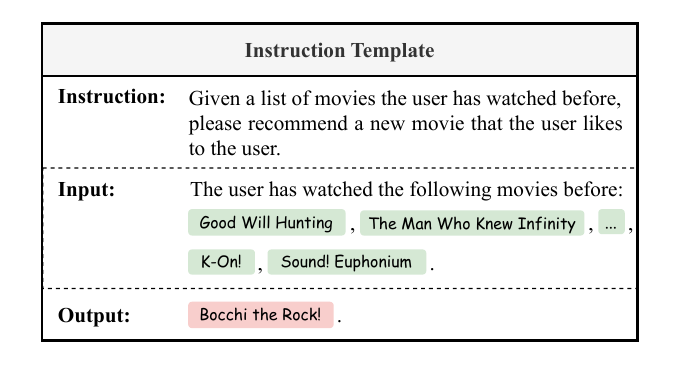}
    \caption{
        Instruction prompt template.
    }
    \Description{Prompt example.}
    \label{fig:prompt}
\end{figure}
% END FIGURE: PROMPT ------------------------------------------------

%%%%%%%%%%%%%%%%%%%%%%%%%%%%%%%%%%%%%%%%%%%%%%%%%%%%%%%%%%%%%%%%%%%%%
% PRELIMINARIES -----------------------------------------------------
\section{Preliminaries} \label{sec:preliminaries}

In this section, we provide the necessary preliminaries for understanding our work. We formulate the task of LLM-based RS in \cref{sec:preliminaries:task-formulation}, introduce beam search in \cref{sec:preliminaries:beam-search}, and discuss the limitations of SFT due to the training-inference inconsistency in \cref{sec:preliminaries:limitations-of-sft}.

%%%%%%%%%%%%%%%%%%%%%%%%%%%%%%%%%%%%%%%%%%%%%%%%%%%%%%%%%%%%%%%%%%%%%
% PRELIMINARIES: TASK FORMULATION -----------------------------------
\subsection{Task Formulation} \label{sec:preliminaries:task-formulation}

In this work, we focus on sequential recommendation~\citep{hidasi2015session,tang2018personalized,kang2018self,chen2023bias,lin2025recommendation,wang2024distributionally,chen2021autodebias}, a widely adopted recommendation scenario that aims to model the temporal dynamics of user preferences. Formally, given a recommender system with the corresponding user set $\mathcal{U}$ and the item set $\mathcal{I}$, each user $u \in \mathcal{U}$ is associated with a historical interaction sequence $H_u = [h_1, h_2, \cdots, h_{N - 1}]$, where $h_n \in \mathcal{I}$ is the $n$-th interacted item. The goal of sequential recommendation is to predict the next item $h_N$ that a user would likely interact with based on the previous interactions $H_u$.

Recently, given the impressive open-domain knowledge and semantic reasoning capabilities of Large Language Models (LLMs)~\citep{achiam2023gpt,bai2023qwen,touvron2023llama}, there has been a surge of interest in leveraging them for sequential recommendation (\ie LLM-based RS)~\citep{bao2025bi,bao2024decoding,kim2024large,liao2024llara}. In this paradigm, the historical interactions $H_u$ are transformed into a structured prompt $x$ using a predefined template, where each item $h_n \in H_u$ is represented by its textual description (\eg title or metadata), as illustrated in \cref{fig:prompt}. The prompt $x$ is subsequently employed to instruct LLMs to generate the textual description $y$ of the predicted item, where the generation is driven by the probability $P_\theta(y | x)$ estimated by LLMs with parameters $\theta$.

Notably, LLMs operate at a fine-grained semantic token level, sequentially outputting the tokens of the predicted item $y$ according to the token-wise conditional probability $P_\theta(y_t | y_{< t}, x)$, where $y_t$ denotes the $t$-th token in $y$, and $y_{< t}$ denotes the prefix tokens preceding $y_t$. Consequently, the overall probability of the predicted item $y$ can be factorized as the product of token probabilities, \ie $P_\theta(y | x) = \prod_{t = 1}^{|y|} P_\theta(y_t | y_{< t}, x)$, where $|y|$ is the length of $y$. Such fine-grained token-level semantic modeling empowers LLMs to capture nuanced user preferences, offering a promising avenue for advancing recommendation performance.

% END: PRELIMINARIES: TASK FORMULATION ------------------------------
%%%%%%%%%%%%%%%%%%%%%%%%%%%%%%%%%%%%%%%%%%%%%%%%%%%%%%%%%%%%%%%%%%%%%

% FIGURE: PRUNING RATE AND NDCG COMPARISON --------------------------
% END FIGURE: PRUNING RATE AND NDCG COMPARISON ----------------------

%%%%%%%%%%%%%%%%%%%%%%%%%%%%%%%%%%%%%%%%%%%%%%%%%%%%%%%%%%%%%%%%%%%%%
% PRELIMINARIES: BEAM SEARCH ----------------------------------------
\subsection{Beam Search} \label{sec:preliminaries:beam-search}

During the inference stage, LLM-based RS aim to retrieve the top-$K$ items with the highest probabilities $P_\theta(y | x)$. However, given the vast item space and significant computational cost of LLMs, exhaustively enumerating all items and calculating their probabilities is prohibitively expensive. To address this challenge, an efficient approximate decoding strategy, known as \emph{beam search}, is widely adopted in LLM-based RS~\citep{bao2024decoding,lin2025igd,wang2025msl}. This strategy leverages the generative nature of LLMs and employs a greedy search principle, achieving a balance between accuracy and efficiency.

\noindentparnoline{Beam Search Procedure.}
Rather than directly assessing the probabilities $P_\theta(y | x)$ of all items, beam search employs an iterative token-wise extension mechanism. Specifically, at each decoding step, beam search progressively extends the sequences (\aka beams) to their next tokens, and retains only a fixed number (\ie the beam width $B$) of the most promising candidate sequences, while pruning the rest. The detailed procedure is as follows (also illustrated in \cref{fig:intro}):

\begin{itemize}[topsep=3pt,leftmargin=10pt,itemsep=0pt]
    \item \emph{Initialization}: at the initial step $t = 0$, initialize the candidate set $\mathcal{B}_{< 1} = \{b_{< 1}\}$ with a single candidate $b_{< 1} = \texttt{"<BOS>"}$, where \texttt{<BOS>} is a special token denoting the beginning of the sequence.
    \item \emph{Expansion}: at step $t \geq 1$, for each candidate sequence $b_{< t}$ from the previous step's candidate set $\mathcal{B}_{< t}$, extend it with all possible next tokens $b_t$ to form the extended candidates $b_{< t + 1}$. The probabilities of the extended candidates are computed as $P_\theta(b_{< t + 1} | x) = P_\theta(b_{< t} | x) \cdot P_\theta(b_t | b_{< t}, x)$.
    \item \emph{Pruning}: sort all extended candidates $b_{< t + 1}$ by their probabilities $P_\theta(b_{< t + 1} | x)$, and retain only the top-$B$ candidates with the highest probabilities as the new candidate set $\mathcal{B}_{< t + 1}$ for the next step. All other candidates are pruned.
    \item \emph{Termination}: repeat the expansion and pruning steps until all candidates are completed by generating a special end-of-sequence token \texttt{<EOS>} or reaching the predefined maximum length. This results in a final candidate set with $B$ completed sequences, from which the top-$K$ items with the highest overall probabilities are selected as the final recommendations.
\end{itemize}

\noindentparnoline{Suboptimal Pruning in Beam Search.}
Despite its efficiency, beam search is inherently a locally greedy decoding strategy and is therefore not guaranteed to yield the optimal retrieval results~\citep{freitag2017beam}. Specifically, at each decoding step, it retains only the top-$B$ candidates $b_{< t}$ with the highest \emph{prefix probabilities} $P_\theta(b_{< t} | x)$, potentially pruning candidates that could lead to higher \emph{overall probabilities} $P_\theta(y | x)$ after generating subsequent tokens. For example, as illustrated in \cref{fig:intro}(a), at the first decoding step, the candidate sequence "\textcolor{beamsearchred}{Bocchi}" is pruned because its prefix probability (25\%) is lower than that of the other two candidates, \ie "\textcolor{beamsearchblue}{The}" (30\%) and "\textcolor{beamsearchblue}{A}" (45\%). However, one of the complete items starting with this prefix, \ie "\textcolor{beamsearchred}{Bocchi the Rock!}", indeed has the highest overall probability (23\%) among all items. In this case, positive items with high probability are incorrectly pruned early due to the greedy nature of beam search, which can significantly hinder retrieval accuracy and lead to low-relevance recommendations~\citep{vijayakumar2018diverse,zeng2024planning}.

% END: PRELIMINARIES: BEAM SEARCH -----------------------------------
%%%%%%%%%%%%%%%%%%%%%%%%%%%%%%%%%%%%%%%%%%%%%%%%%%%%%%%%%%%%%%%%%%%%%

% TABLE: PR@K AND NPR@K ---------------------------------------------
\begin{figure}[t]
    \centering
    \includegraphics[width=\columnwidth]{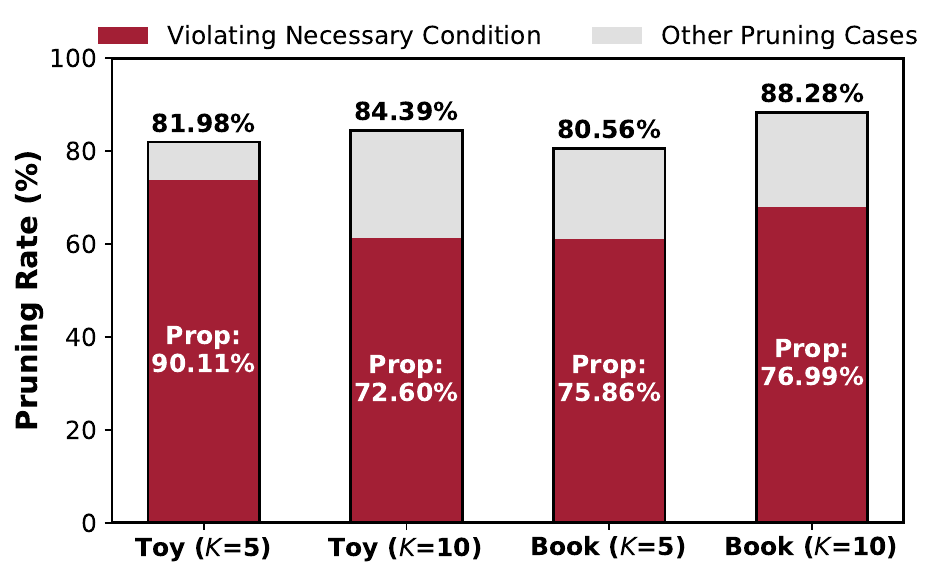}
    \caption{
        An overwhelming majority of positive items with high overall probabilities (\ie top-$K$ ranked among candidates) are pruned during beam search, where failure to meet the necessary condition \eqnref{eq:necessary_condition} is the primary cause (marked as \textcolor{pruning_red}{red} rectangles). Here, we set beam width $B = 10$ and number of recommended items $K \in \{5, 10\}$. "Prop" denotes the proportion of pruning cases attributable to necessary condition violations (\textcolor{pruning_red}{red}), consistently accounting for over 70\% of all pruning cases (\textcolor{pruning_red}{red} and \textcolor{pruning_gray}{gray}) across all settings.
    }
    \Description{Failure to meet the necessary condition is the primary cause for incorrect pruning of positive items during beam search.}
    \label{fig:pr_and_tpr_pruning_rate}
\end{figure}
% END TABLE: PR@K AND NPR@K -----------------------------------------

%%%%%%%%%%%%%%%%%%%%%%%%%%%%%%%%%%%%%%%%%%%%%%%%%%%%%%%%%%%%%%%%%%%%%
% PRELIMINARIES: LIMITATIONS OF SFT ---------------------------------

\subsection{Limitations of SFT} \label{sec:preliminaries:limitations-of-sft}

To adapt LLMs to recommendation scenarios, existing LLM-based RS typically employ supervised fine-tuning (SFT)~\citep{bao2024decoding,lin2025igd,lu2025dual,wang2025msl}. In this process, the recommendation data is organized into a training dataset consisting of multiple prompt-response pairs $(x, y)$ as training instances, where $x$ is the constructed prompt and $y$ is the textual description of the positive item (\cf \cref{fig:prompt}). Subsequently, the LLM is adapted by optimizing the following SFT objective:
\begin{equation} \label{eq:sft}
    \min_\theta \mathcal{L}_{\textnormal{SFT}}(x, y; \theta) = - \log P_\theta(y | x) = - \sum_{t = 1}^{|y|} \log P_\theta(y_t | y_{< t}, x).
\end{equation}
By minimizing this loss, the model is guided to maximize the overall probability $P_\theta(y | x)$ of the positive item $y$ conditioned on the prompt $x$, thereby aligning the LLM with recommendation tasks. More importantly, the SFT objective can be decomposed into a sum of token-wise cross-entropy losses (\ie $-\log P_\theta(y_t | y_{< t}, x)$), which can be efficiently optimized through one forward pass of the LLM per training instance, facilitating scalable training.

\noindentparnoline{Training-Inference Inconsistency.}
While SFT has shown promising results, this conventional strategy could be suboptimal for LLM-based RS due to the \emph{training-inference inconsistency} between SFT learning and beam search decoding. Specifically, the SFT objective aims to maximize the overall probability of positive items. However, \textbf{\emph{during beam search, a high overall probability does not necessarily ensure successful retrieval}}. As discussed in \cref{sec:preliminaries:beam-search}, the greedy pruning behavior of beam search can easily eliminate positive items whose prefixes do not exhibit sufficiently high local probability, regardless of their global high probability. As exemplified in \cref{fig:intro}(a), the positive item "Bocchi the Rock!" is pruned due to the relatively low probability of its prefix "Bocchi", despite having the highest overall probability among all items. Critically, this situation is alarmingly common rather than exceptional. Our empirical analysis, as exhibited in \cref{fig:pr_and_tpr_pruning_rate,fig:pruning-rate}, reveals a striking finding: \textbf{\emph{among positive items with high overall probabilities, an overwhelming majority (\eg 88.28\% in Book and 84.39\% in Toy datasets) are pruned by beam search and consequently fail to appear in the final recommendations}}. This remarkable percentage underscores that the training-inference inconsistency represents a fundamental bottleneck in LLM-based RS, severely compromising their effectiveness and demanding immediate attention. Thus, this work explores a novel beam-search-aware fine-tuning approach to mitigate this critical inconsistency.

% END: PRELIMINARIES: TRAINING-INFERENCE GAP ------------------------
%%%%%%%%%%%%%%%%%%%%%%%%%%%%%%%%%%%%%%%%%%%%%%%%%%%%%%%%%%%%%%%%%%%%%

% END: PRELIMINARIES ------------------------------------------------
%%%%%%%%%%%%%%%%%%%%%%%%%%%%%%%%%%%%%%%%%%%%%%%%%%%%%%%%%%%%%%%%%%%%%

% FIGURE: RUNNING TIME COMPARISON -----------------------------------
\begin{figure}[t]
    \centering
    \includegraphics[width=0.98\columnwidth]{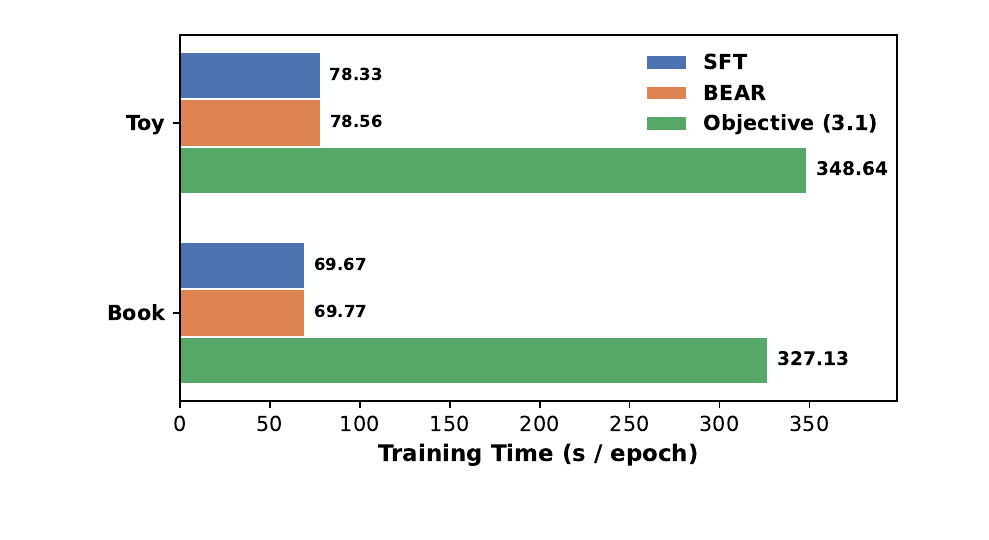}
    \caption{
        Running time comparison of three training objectives: SFT, BEAR, and objective \eqnref{eq:sufficient_condition}. While SFT and BEAR have comparable training time, optimizing the objective \eqnref{eq:sufficient_condition} is significantly more time-consuming due to the need for beam search simulation during training. Refer to \cref{fig:training-efficiency} for complete results in training efficiency.
    }
    \Description{Running time comparison of three training objectives.}
    \label{fig:runtime}
\end{figure}
% END FIGURE: RUNNING TIME COMPARISON -------------------------------

%%%%%%%%%%%%%%%%%%%%%%%%%%%%%%%%%%%%%%%%%%%%%%%%%%%%%%%%%%%%%%%%%%%%%
% METHODOLOGY -------------------------------------------------------
\section{Methodology} \label{sec:methodology}

To bridge the training-inference inconsistency, we present \textbf{BEAR} (\textbf{\uline{B}}eam-S\textbf{\uline{E}}arch-\textbf{\uline{A}}ware \textbf{\uline{R}}egularization), a novel \emph{beam-search-aware} objective that explicitly considers the impact of beam search. We first present a straightforward yet time-costly solution in \cref{sec:methodology:sufficient}, then introduce our BEAR objective in \cref{sec:methodology:bear}.

%%%%%%%%%%%%%%%%%%%%%%%%%%%%%%%%%%%%%%%%%%%%%%%%%%%%%%%%%%%%%%%%%%%%%
% METHODOLOGY: SUFFICIENT CONDITION ---------------------------------
\subsection{A Straightforward Solution} \label{sec:methodology:sufficient}

In \cref{sec:preliminaries:limitations-of-sft}, we have identified the limitations of SFT due to its beam search unawareness. Since SFT merely maximizes the overall probability of the positive item, it cannot prevent low-probability prefixes from being pruned during beam search inference, which may lead to degraded recommendation performance.

\noindentparnoline{Straightforward Objective.}
To address this issue, a straightforward idea is to directly optimize the objective for retaining the positive item throughout the beam search process, \ie ensuring that every prefix $y_{< t}$ of the positive item $y$ is always ranked within the top-$B$ candidates at each beam search step $t$, where $B$ is the beam width. Formally, this requirement can be expressed as:
\begin{equation} \label{eq:sufficient_condition}
    \max_\theta \prod_{t = 1}^{|y| + 1} \mathbb{I} \left( P_\theta(y_{< t} | x) \geq P_\theta(b_{< t}^{B} | x) \right),
\end{equation}
where $b_{< t}^{B}$ denotes the beam search candidate ranked exactly in the $B$-th position at decoding step $t - 1$ (\cf \cref{sec:preliminaries:beam-search}), and $\mathbb{I}(\cdot)$ is the indicator function. The right-hand term $P_\theta(b_{< t}^{B} | x)$ can be understood as a threshold --- if the probability of the prefix $y_{< t}$, \ie the left-hand term $P_\theta(y_{< t} | x)$, exceeds this threshold $P_\theta(b_{< t}^{B} | x)$, the prefix will be preserved by beam search; otherwise, it will be pruned immediately. Optimizing the objective \eqnref{eq:sufficient_condition} systematically elevates the probability of each positive prefix above the corresponding threshold, thereby reducing the risk that the positive item will be incorrectly pruned during beam search.

\noindentparnoline{Limitations on Computational Efficiency.}
Despite its conceptual simplicity, this straightforward solution is computationally impractical in real-world settings. Specifically, calculating the threshold $P_\theta(b_{< t}^{B} | x)$ in objective \eqnref{eq:sufficient_condition} necessitates simulating the entire beam search procedure during training for each instance $(x, y)$. This involves token-by-token expansion of candidate sequences and requires up to $B$ times additional forward passes of the LLM per training instance compared to SFT. Worse still, the strictly sequential dependency makes this process inherently difficult to parallelize and thus heavily time-consuming. Even with a relatively small beam width (\eg $B = 10$), our empirical analysis shows that optimizing \eqnref{eq:sufficient_condition} incurs over $4.45 \times$ runtime cost compared to basic SFT in practice, making it unsuitable for real-world applications (\cf \cref{fig:runtime}). This motivates the development of an efficient surrogate strategy to approximate the objective while avoiding the full beam search simulation during training.

% END: METHODOLOGY: SUFFICIENT CONDITION ----------------------------
%%%%%%%%%%%%%%%%%%%%%%%%%%%%%%%%%%%%%%%%%%%%%%%%%%%%%%%%%%%%%%%%%%%%%

%%%%%%%%%%%%%%%%%%%%%%%%%%%%%%%%%%%%%%%%%%%%%%%%%%%%%%%%%%%%%%%%%%%%%
% METHODOLOGY: BEAR: BEAM-SEARCH-AWARE REGULARIZATION ---------------
\subsection{BEAR: Beam-Search-Aware Regularization} \label{sec:methodology:bear}

\noindentparnolinenospace{Necessary Condition Objective.}
While directly optimizing the objective \eqnref{eq:sufficient_condition} is challenging, we empirically observe an intriguing phenomenon: \textbf{\emph{the majority of incorrect pruning cases are in fact caused by the violation of a necessary condition}}, as illustrated in \cref{fig:pr_and_tpr_pruning_rate}. This condition stipulates that the token-wise probability $P_\theta(y_t | y_{< t}, x)$ of each positive token $y_t$ must rank within the top-$B$ among all possible next tokens; otherwise, it will be pruned immediately by the beam search, since only the top-$B$ extended candidates are retained at each decoding step (\cf \cref{sec:preliminaries:beam-search}). Formally, this necessary condition can be expressed as:
\begin{equation} \label{eq:necessary_condition}
    \max_\theta \prod_{t = 1}^{|y|} \mathbb{I} \left( P_\theta(y_t | y_{< t}, x) \geq \beta_t^B \right),
\end{equation}
where $\beta_t^B = \operatorname{top-\mathit{B}} \{P_\theta(\cdot | y_{< t}, x)\}$ denotes the $B$-th highest token-wise probability among all possible next tokens at step $t$. In contrast to the original objective \eqnref{eq:sufficient_condition}, which enforces that \emph{every positive prefix} is retained within the top-$B$ beam search candidates, this formulation operates at the granularity of \emph{individual positive tokens}. The goal of objective \eqnref{eq:necessary_condition} is to raise the probability of each positive token $P_\theta(y_t | y_{< t}, x)$ above the token-wise threshold $\beta_t^B$, thereby ensuring that the extended positive prefix $y_{< t+1}$ at least remains within the top-$B$ among all possible continuations extended from the prefix $y_{< t}$, which is a necessary condition for finally retaining this positive prefix in the next step of beam search.

Notably, although this objective is a \emph{weaker constraint} than \eqnref{eq:sufficient_condition}, it remains a strong enough constraint and serves as an \emph{effective and computationally efficient surrogate} for optimization. On the one hand, as shown in \cref{fig:pr_and_tpr_pruning_rate}, failure to satisfy this necessary condition is the \emph{primary} cause of incorrect pruning, accounting for over 70\% of all cases. Therefore, directly optimizing this necessary condition can substantially mitigate the training-inference inconsistency and improve recommendation performance. On the other hand, unlike \eqnref{eq:sufficient_condition}, this objective does not require simulating the beam search procedure and only involves computing the token-wise top-$B$ threshold $\beta_t^B$. Note that the probabilities $P_\theta(\cdot | y_{< t}, x)$ for all valid next tokens of each positive prefix $y_{< t}$ can be efficiently calculated by applying softmax to the output logits of LLMs, while these logits are already computed during SFT. Therefore, calculating this top-$B$ threshold $\beta_t^B$ incurs no additional forward passes of LLMs. As a result, its computational cost is comparable to standard SFT and is significantly more efficient than directly optimizing \eqnref{eq:sufficient_condition}, as evidenced by our empirical results in \cref{fig:runtime,fig:training-efficiency}.

% TABLE: DATASET STATISTICS -----------------------------------------
\begin{table}[t]
    \centering
    \caption{Dataset statistics.}
    \label{tab:statistics}
    \setlength{\tabcolsep}{5pt}
    \begin{tabular}{l|ccccc}
    \toprule
    \textbf{Dataset} & \textbf{\#Users} & \textbf{\#Items} & \textbf{\#Interactions} & \textbf{Density} \\
    \midrule
    \textbf{Office}   
        & 4895   & 2414   & 53149         & 0.4498\%   \\
    \textbf{Book}     
        & 16559  & 6344   & 151928        & 0.1446\%   \\
    \textbf{Toy}      
        & 19124  & 11758  & 165247        & 0.0735\%   \\
    \textbf{Clothing} 
        & 39230  & 22948  & 277534        & 0.0308\%   \\
    \bottomrule
    \end{tabular}
\end{table}
% END TABLE: DATASET STATISTICS -------------------------------------

% TABLE: MAIN EXPERIMENT RESULTS ------------------------------------
\begin{table*}[t]
    \centering
    \caption{
        Overall performance comparison with various baselines. The LLM-based RS methods are applied ("+") on the BIGRec~\citep{bao2025bi} backbone. N@$K$ and H@$K$ denote NDCG@$K$ and HitRatio@$K$ metrics, respectively. The best results are highlighted in bold, and the best baselines are underlined. "\textcolor{darkred}{\textbf{Imp.\%}}" indicates the relative improvement of BEAR over the best baseline.
    }
    \label{tab:main-results}
  \resizebox{\textwidth}{!}{
    \begin{tabular}{l|cccc|cccc|cccc|cccc}
    \toprule
    \multicolumn{1}{c|}{\multirow{2}[4]{*}{\textbf{Methods}}} 
        & \multicolumn{4}{c|}{\textbf{Office}} 
        & \multicolumn{4}{c|}{\textbf{Book}} 
        & \multicolumn{4}{c|}{\textbf{Toy}} 
        & \multicolumn{4}{c}{\textbf{Clothing}} \\
    \cmidrule(lr){2-5} \cmidrule(lr){6-9} \cmidrule(lr){10-13} \cmidrule(lr){14-17}
        & \textbf{N@5} & \textbf{N@10} & \textbf{H@5} & \textbf{H@10} 
        & \textbf{N@5} & \textbf{N@10} & \textbf{H@5} & \textbf{H@10} 
        & \textbf{N@5} & \textbf{N@10} & \textbf{H@5} & \textbf{H@10} 
        & \textbf{N@5} & \textbf{N@10} & \textbf{H@5} & \textbf{H@10} \\
    \midrule
    SASRec      & 0.0127 & 0.0196 & 0.0223 & 0.0447 & 0.0061 & 0.0080 & 0.0100 & 0.0157 & 0.0089 & 0.0109 & 0.0148 & 0.0211 & 0.0022 & 0.0033 & 0.0036 & 0.0072 \\
    BERT4Rec    & 0.0164 & 0.0206 & 0.0264 & 0.0396 & 0.0100 & 0.0121 & \uline{0.0158} & 0.0223 & 0.0101 & 0.0118 & 0.0137 & 0.0188 & 0.0028 & 0.0034 & 0.0033 & 0.0053 \\
    DROS        & 0.0195 & 0.0240 & 0.0265 & 0.0405 & 0.0072 & 0.0096 & 0.0116 & 0.0192 & 0.0105 & 0.0126 & 0.0148 & 0.0213 & 0.0038 & 0.0046 & 0.0048 & 0.0074 \\
    \midrule
    LLM-CF      & 0.0161 & 0.0200 & 0.0306 & 0.0426 & 0.0077 & 0.0100 & 0.0141 & 0.0212 & 0.0077 & 0.0101 & 0.0148 & 0.0223 & 0.0022 & 0.0034 & 0.0042 & 0.0078 \\
    DLLM2Rec    & 0.0166 & 0.0231 & 0.0322 & 0.0530 & 0.0068 & 0.0094 & 0.0112 & 0.0194 & 0.0080 & 0.0110 & 0.0156 & 0.0248 & 0.0028 & 0.0045 & 0.0056 & 0.0106 \\
    \midrule
    BIGRec      & 0.0105 & 0.0226 & 0.0213 & 0.0587 & 0.0068 & 0.0101 & 0.0128 & 0.0228 & 0.0122 & 0.0157 & 0.0200 & 0.0307 & 0.0043 & 0.0071 & 0.0086 & 0.0172 \\
    + D$^3$      & 0.0142 & 0.0237 & 0.0291 & 0.0592 & 0.0057 & 0.0093 & 0.0112 & 0.0221 & 0.0163 & 0.0202 & 0.0257 & 0.0375 & 0.0045 & 0.0072 & 0.0088 & 0.0172 \\
    + CFT        & 0.0183 & 0.0272 & 0.0306 & 0.0582 & 0.0051 & 0.0090 & 0.0107 & 0.0224 & 0.0138 & 0.0173 & 0.0217 & 0.0325 & 0.0045 & 0.0076 & 0.0086 & 0.0182 \\
    + D$^2$LR    & 0.0243 & 0.0311 & 0.0395 & 0.0603 & 0.0070 & 0.0089 & 0.0128 & 0.0183 & 0.0142 & 0.0177 & 0.0227 & 0.0338 & 0.0051 & 0.0081 & 0.0098 & 0.0194 \\
    + IGD        & 0.0187 & 0.0266 & 0.0312 & 0.0566 & 0.0061 & 0.0095 & 0.0125 & 0.0228 & 0.0191 & 0.0239 & 0.0273 & 0.0421 & 0.0059 & 0.0084 & 0.0098 & 0.0178 \\
    + MSL        & \uline{0.0346} & \uline{0.0388} & \uline{0.0504} & 0.0634 & 0.0087 & 0.0117 & 0.0137 & 0.0228 & \uline{0.0239} & \uline{0.0301} & \uline{0.0359} & \uline{0.0551} & \uline{0.0075} & \uline{0.0115} & 0.0136 & 0.0260 \\
    % + DPO        & 0.0247 & 0.0323 & 0.0400 & 0.0634 & 0.0100 & 0.0117 & 0.0139 & 0.0196 & 0.0165 & 0.0216 & 0.0263 & 0.0419 & 0.0054 & 0.0087 & 0.0104 & 0.0206 \\
    + S-DPO      & 0.0209 & 0.0292 & 0.0374 & 0.0629 & 0.0082 & 0.0102 & 0.0125 & 0.0189 & 0.0181 & 0.0227 & 0.0288 & 0.0432 & 0.0054 & 0.0093 & 0.0110 & 0.0232 \\
    + RosePO     & 0.0266 & 0.0338 & 0.0436 & \uline{0.0660} & \uline{0.0117} & \uline{0.0127} & 0.0153 & 0.0185 & 0.0114 & 0.0174 & 0.0194 & 0.0378 & 0.0066 & 0.0110 & \uline{0.0138} & \uline{0.0272} \\
    + SPRec      & 0.0249 & 0.0326 & 0.0416 & 0.0655 & 0.0072 & 0.0110 & 0.0130 & \uline{0.0244} & 0.0148 & 0.0175 & 0.0234 & 0.0317 & 0.0047 & 0.0069 & 0.0082 & 0.0152 \\
    \midrule
    \textbf{+ BEAR} & \textbf{0.0365} & \textbf{0.0421} & \textbf{0.0545} & \textbf{0.0717} & \textbf{0.0146} & \textbf{0.0177} & \textbf{0.0203} & \textbf{0.0297} & \textbf{0.0254} & \textbf{0.0316} & \textbf{0.0380} & \textbf{0.0574} & \textbf{0.0082} & \textbf{0.0127} & \textbf{0.0148} & \textbf{0.0290} \\
    \textcolor{darkred}{\textbf{Imp.\%}} & \textcolor{darkred}{\textbf{+5.49\%}} & \textcolor{darkred}{\textbf{+8.51\%}} & \textcolor{darkred}{\textbf{+8.13\%}} & \textcolor{darkred}{\textbf{+8.64\%}} 
        & \textcolor{darkred}{\textbf{+24.79\%}} & \textcolor{darkred}{\textbf{+39.37\%}} & \textcolor{darkred}{\textbf{+28.48\%}} & \textcolor{darkred}{\textbf{+21.72\%}} 
        & \textcolor{darkred}{\textbf{+6.28\%}} & \textcolor{darkred}{\textbf{+4.98\%}} & \textcolor{darkred}{\textbf{+5.85\%}} & \textcolor{darkred}{\textbf{+4.17\%}} 
        & \textcolor{darkred}{\textbf{+9.33\%}} & \textcolor{darkred}{\textbf{+10.43\%}} & \textcolor{darkred}{\textbf{+7.25\%}} & \textcolor{darkred}{\textbf{+6.62\%}} \\
    \bottomrule
    \end{tabular}
  }
\end{table*}
% END TABLE: MAIN EXPERIMENT RESULTS --------------------------------

\noindentparnoline{Beam-Search-Aware Regularization (BEAR).}
Despite the potential benefits of optimizing the necessary condition \eqnref{eq:necessary_condition}, this approach poses significant challenges due to its inherent \emph{discontinuity}, which hinders gradient-based optimization. To tackle these issues, we follow the principle of Empirical Risk Minimization (ERM)~\citep{vapnik2013nature} and formulate the risk indicating the violation of this necessary condition as:
% This assumption is reasonable, since the output logits of LLMs are typically bounded, making the token probabilities calculated by softmax on the logits also bounded.
\begin{equation} \label{eq:risk_function}
    \mathcal{R}_t(x, y; \theta) = \log \mathbb{I} \left( \log \beta_t^B - \log P_\theta(y_t | y_{< t}, x) > 0 \right).
\end{equation}
Here, the $\Delta_t^B = \log \beta_t^B - \log P_\theta(y_t | y_{< t}, x)$ term\footnote{In objective \eqnref{eq:necessary_condition}, the indicator function $\mathbb{I}\left(P_\theta(y_t | y_{< t}, x) \geq \beta_t^B\right)$ is equivalent to $\mathbb{I}\left(\log \beta_t^B - \log P_\theta(y_t | y_{< t}, x) \leq 0\right)$, which operates on the difference of log-probabilities and thus facilitates optimization.} can be interpreted as the \emph{pruning margin} at step $t$. A positive pruning margin indicates that the token-wise probability of $y_t$ falls below the top-$B$ threshold $\beta_t^B$, which corresponds to the violation of the necessary condition \eqnref{eq:necessary_condition} and leads to immediate pruning at step $t$. Therefore, minimizing this risk for each step is fundamentally equivalent to optimizing the necessary condition for beam search survival.

To enable gradient-based optimization and effectively minimize the risk \eqnref{eq:risk_function}, we surrogate the indicator function $\mathbb{I}(\Delta_t^B > 0)$ with the sigmoid function $\sigma_{\xi}(\Delta_t^B) = 1 / (1 + \exp(-\Delta_t^B / \xi))$, where $\xi$ is a temperature hyperparameter that controls the smoothness\footnote{As the temperature $\xi \to 0$, the sigmoid function $\sigma_{\xi}(\Delta_t^B)$ converges to $\mathbb{I}(\Delta_t^B > 0)$.}~\citep{yang2024psl,yang2025breaking,zhang2026talos}. This surrogate provides a smooth and tight approximation to the discontinuous indicator function, resulting in the following differentiable \textbf{\emph{beam-search-aware regularization}} term\footnote{In objectives \eqnref{eq:risk_function} and \eqnref{eq:beam_aware_regularization}, we take the logarithm to convert maximization of the product into minimization of the sum of risks, facilitating token-wise optimization. For mathematically rigorousness, we assume that the token probabilities are lower bounded by a small positive constant, ensuring the well-definedness of the logarithm operation~\citep{wu2024effectiveness,yang2024psl,yang2025breaking,zhang2025advancing}.}:
\begin{equation} \label{eq:beam_aware_regularization}
    \mathcal{L}_{\textnormal{reg}}(x, y; \theta)
    = \sum_{t = 1}^{|y|} \log \sigma_{\xi} \left(\log \beta_t^B - \log P_\theta(y_t | y_{< t}, x) \right).
\end{equation}
By minimizing this regularization term, the model is guided to minimize the pruning margin $\Delta_t^B$ at each step, thereby effectively reducing the risk of incorrect pruning during beam search. Notably, while the standard SFT objective \eqnref{eq:sft} also increases the probabilities of positive tokens, it treats all tokens equally without considering the beam search dynamics, leading to massive premature pruning and noteworthy training-inference inconsistency (\cf \cref{sec:preliminaries:limitations-of-sft}). In contrast, regularization \eqnref{eq:beam_aware_regularization} explicitly accounts for the pruning margin, dynamically evaluating the probabilities of positive tokens $P_\theta(y_t | y_{< t}, x)$ relative to their respective top-$B$ thresholds $\beta_t^B$, thus effectively mitigating the training-inference inconsistency.

\begin{figure*}[t]
    \centering
    \includegraphics[width=\textwidth]{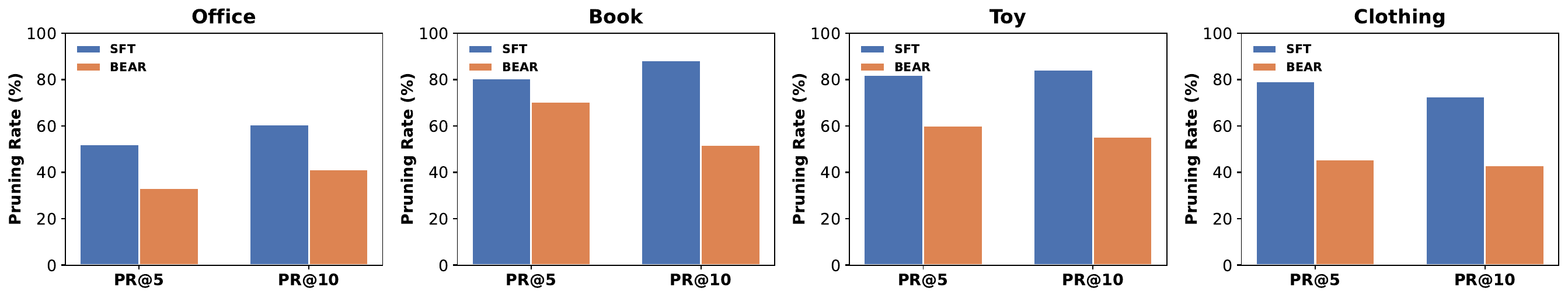}
    \caption{
        Pruning Rate (PR@$K$) comparison between SFT and BEAR. During beam search with beam size $B$, PR@$K$ measures the proportion of incorrectly pruned positive items with high overall probabilities (\ie top-$K$ ranked among candidates). A lower PR@$K$ indicates better training-inference alignment. Results show that BEAR significantly reduces PR@$K$ compared to SFT.
    }
    \Description{Pruning Rate Comparison between SFT and BEAR.}
    \label{fig:pruning-rate}
\end{figure*}

Finally, by incorporating the regularization term \eqnref{eq:beam_aware_regularization} into the SFT objective \eqnref{eq:sft}, we obtain the proposed \textbf{BEAR objective}:
\begin{equation} \label{eq:bear}
    \min_\theta \mathcal{L}_{\textnormal{BEAR}}(x, y; \theta)
    = \mathcal{L}_{\textnormal{SFT}}(x, y; \theta) + \lambda \mathcal{L}_{\textnormal{reg}}(x, y; \theta),
\end{equation}
where $\lambda > 0$ is a weight hyperparameter that controls the regularization strength. By incorporating beam-search awareness with SFT, BEAR not only improves the overall probability of positive items (\cf \cref{tab:main-results}), but also effectively reduces the risk of incorrectly pruning positive items during beam search (\cf \cref{fig:pruning-rate}). These effects collectively alleviate the training-inference inconsistency and enhance recommendation performance, leading to an effective and efficient fine-tuning objective for LLM-based RS.
\begin{table*}[t]
  \centering
  \caption{
    Performance comparison using different recommendation backbones, including LLaRA~\citep{liao2024llara} and A-LLMRec~\citep{kim2024large}. We adopt MSL~\citep{wang2025msl} as the representative SFT method. N@$K$ and H@$K$ denote NDCG@$K$ and HitRatio@$K$ metrics, respectively. The best results are highlighted in bold. "\textcolor{darkred}{\textbf{Imp.\%}}" indicates the relative improvement of BEAR over the best baseline.
  }
  \resizebox{\textwidth}{!}{
    \begin{tabular}{l|cccc|cccc|cccc|cccc}
    \toprule
    \multicolumn{1}{c|}{\multirow{2}[4]{*}{\textbf{Methods}}} 
        & \multicolumn{4}{c|}{\textbf{Office}} 
        & \multicolumn{4}{c|}{\textbf{Book}} 
        & \multicolumn{4}{c|}{\textbf{Toy}} 
        & \multicolumn{4}{c}{\textbf{Clothing}} \\
    \cmidrule(lr){2-5} \cmidrule(lr){6-9} \cmidrule(lr){10-13} \cmidrule(lr){14-17}     
        & \textbf{N@5} & \textbf{N@10} & \textbf{H@5} & \textbf{H@10} 
        & \textbf{N@5} & \textbf{N@10} & \textbf{H@5} & \textbf{H@10} 
        & \textbf{N@5} & \textbf{N@10} & \textbf{H@5} & \textbf{H@10} 
        & \textbf{N@5} & \textbf{N@10} & \textbf{H@5} & \textbf{H@10} \\
    \midrule
    % --- Backbone 1: LLaRA ---
    LLaRA       & 0.0110 & 0.0234 & 0.0234 & 0.0618 & 0.0068 & 0.0113 & 0.0135 & 0.0267 & 0.0129 & 0.0163 & 0.0204 & 0.0309 & 0.0036 & 0.0071 & 0.0074 & 0.0184 \\
    + MSL       & 0.0380 & 0.0423 & 0.0561 & 0.0696 & 0.0078 & 0.0105 & 0.0132 & 0.0217 & 0.0230 & 0.0290 & 0.0346 & 0.0534 & 0.0082 & 0.0116 & 0.0146 & 0.0254 \\
    % \midrule
    \textbf{+ BEAR} & \textbf{0.0407} & \textbf{0.0457} & \textbf{0.0592} & \textbf{0.0748} 
        & \textbf{0.0123} & \textbf{0.0157} & \textbf{0.0212} & \textbf{0.0315} 
        & \textbf{0.0259} & \textbf{0.0315} & \textbf{0.0386} & \textbf{0.0559} 
        & \textbf{0.0089} & \textbf{0.0129} & \textbf{0.0156} & \textbf{0.0284} \\
    \textcolor{darkred}{\textbf{Imp.\%}} & \textcolor{darkred}{\textbf{+7.11\%}} & \textcolor{darkred}{\textbf{+8.04\%}} & \textcolor{darkred}{\textbf{+5.53\%}} & \textcolor{darkred}{\textbf{+7.47\%}}
        & \textcolor{darkred}{\textbf{+57.69\%}} & \textcolor{darkred}{\textbf{+38.94\%}} & \textcolor{darkred}{\textbf{+57.04\%}} & \textcolor{darkred}{\textbf{+17.98\%}}
        & \textcolor{darkred}{\textbf{+12.61\%}} & \textcolor{darkred}{\textbf{+8.62\%}} & \textcolor{darkred}{\textbf{+11.56\%}} & \textcolor{darkred}{\textbf{+4.68\%}}
        & \textcolor{darkred}{\textbf{+8.54\%}} & \textcolor{darkred}{\textbf{+11.21\%}} & \textcolor{darkred}{\textbf{+6.85\%}} & \textcolor{darkred}{\textbf{+11.81\%}} \\
    \midrule
    % --- Backbone 2: A-LLMRec ---
    A-LLMRec    & 0.0125 & 0.0247 & 0.0265 & 0.0639 & 0.0077 & 0.0111 & 0.0153 & 0.0256 & 0.0126 & 0.0160 & 0.0198 & 0.0305 & 0.0036 & 0.0068 & 0.0072 & 0.0172 \\
    + MSL       & 0.0332 & 0.0373 & 0.0525 & 0.0655 & 0.0094 & 0.0118 & 0.0151 & 0.0224 & 0.0247 & 0.0302 & 0.0361 & 0.0532 & 0.0081 & 0.0124 & 0.0140 & 0.0276 \\
    % \midrule
    \textbf{+ BEAR} & \textbf{0.0392} & \textbf{0.0444} & \textbf{0.0561} & \textbf{0.0722} 
        & \textbf{0.0106} & \textbf{0.0149} & \textbf{0.0189} & \textbf{0.0317} 
        & \textbf{0.0255} & \textbf{0.0312} & \textbf{0.0373} & \textbf{0.0549} 
        & \textbf{0.0087} & \textbf{0.0128} & \textbf{0.0164} & \textbf{0.0288} \\
    \textcolor{darkred}{\textbf{Imp.\%}} & \textcolor{darkred}{\textbf{+18.07\%}} & \textcolor{darkred}{\textbf{+19.03\%}} & \textcolor{darkred}{\textbf{+6.86\%}} & \textcolor{darkred}{\textbf{+10.23\%}}
        & \textcolor{darkred}{\textbf{+12.77\%}} & \textcolor{darkred}{\textbf{+26.27\%}} & \textcolor{darkred}{\textbf{+23.53\%}} & \textcolor{darkred}{\textbf{+23.83\%}}
        & \textcolor{darkred}{\textbf{+3.24\%}} & \textcolor{darkred}{\textbf{+3.31\%}} & \textcolor{darkred}{\textbf{+3.32\%}} & \textcolor{darkred}{\textbf{+3.20\%}}
        & \textcolor{darkred}{\textbf{+7.41\%}} & \textcolor{darkred}{\textbf{+3.23\%}} & \textcolor{darkred}{\textbf{+17.14\%}} & \textcolor{darkred}{\textbf{+4.35\%}} \\
    \bottomrule
    \end{tabular}
  }
  \label{tab:different-backbones}
\end{table*}
% END TABLE: DIFFERENT BACKBONES ------------------------------------

% FIGURE: PRUNING RATE ----------------------------------------------
% END FIGURE: PRUNING RATE ------------------------------------------

%%%%%%%%%%%%%%%%%%%%%%%%%%%%%%%%%%%%%%%%%%%%%%%%%%%%%%%%%%%%%%%%%%%%%
% EXPERIMENTS -------------------------------------------------------
\section{Experiments} \label{sec:experiments}

In this section, we conduct extensive experiments to evaluate the effectiveness of the proposed BEAR objective. Our aim is to answer the following research questions (RQs):
\begin{itemize}[topsep=3pt,leftmargin=10pt,itemsep=0pt]
    \item \textbf{RQ1}: How does BEAR perform in recommendation tasks compared to existing methods?
    \item \textbf{RQ2}: How does BEAR mitigate the training-inference inconsistency and alleviate the incorrect pruning during beam search?
    \item \textbf{RQ3}: How does BEAR generalize to different LLM-based RS backbones, LLM sizes, and fine-tuning objectives?
    \item \textbf{RQ4}: How does the computational efficiency of BEAR compare to other methods?
    \item \textbf{RQ5}: How do the hyperparameters $\lambda$ and $\xi$ of BEAR affect its performance?
\end{itemize}

%%%%%%%%%%%%%%%%%%%%%%%%%%%%%%%%%%%%%%%%%%%%%%%%%%%%%%%%%%%%%%%%%%%%%
% EXPERIMENTS: EXPERIMENTAL SETUP -----------------------------------
\subsection{Experimental Setup} \label{sec:experiments:setup}

\noindentparnolinenospace{Datasets.}
To ensure fair comparison with existing methods, we conduct experiments on four widely-used Amazon datasets~\citep{he2016ups,mcauley2015image}: Office, Book, Toy, and Clothing. The dataset preprocessing conforms to the standard procedure in previous studies~\citep{bao2025bi,wang2025msl}, where the users and items with fewer than 5 interactions are filtered out (\ie 5-core), and a sliding window of size 11 is applied to segment the user interaction sequences. The resulting sequences are then split into training, validation, and test sets with a ratio of 8:1:1 based on the chronological order. The dataset statistics are summarized in \cref{tab:statistics}.

\noindentparnoline{Compared Methods.}
We compare BEAR with the following representative methods:

\begin{itemize}[topsep=3pt,leftmargin=10pt,itemsep=0pt]
    \item \textbf{Conventional RS}: traditional methods for sequential recommendation, including \uline{SASRec} (ICDM '18)~\citep{kang2018self}, \uline{BERT4Rec} (CIKM '19)~\citep{sun2019bert4rec}, and \uline{DROS} (SIGIR '23)~\citep{yang2023generic}.
    \item \textbf{LLM-enhanced RS}: conventional RS models augmented with LLMs as semantic encoder or knowledge enhancer, including \uline{LLM-CF} (CIKM '24)~\citep{sun2024large} and \uline{DLLM2Rec} (RecSys '24)~\citep{cui2024distillation}.
    \item \textbf{LLM-based RS Backbones}: LLMs directly employed for recommendation. Typical backbones include \uline{BIGRec} (TORS '25)~\citep{bao2025bi}, \uline{LLaRA} (SIGIR '24)~\citep{liao2024llara}, and \uline{A-LLMRec} (KDD '24)~\citep{kim2024large}.
    \item \textbf{SFT Methods for LLM-based RS}: various SFT optimization methods applied on LLM-based RS, including \uline{D$^3$} (EMNLP '24)~\citep{bao2024decoding}, \uline{CFT} (CoRR '24)~\citep{zhang2024causality}, \uline{D$^2$LR} (SIGIR '25)~\citep{lu2025dual}, \uline{IGD} (CoRR '25)~\citep{lin2025igd}, and the state-of-the-art method \uline{MSL} (SIGIR '25)~\citep{wang2025msl}.
    % \item \textbf{DPO Methods for LLM-based RS}: direct preference optimization methods that model pairwise preferences for LLM-based RS, including \uline{DPO} (NeurIPS '23)~\citep{rafailov2023direct}, \uline{S-DPO} (NeurIPS '24)~\citep{chen2024softmax}, \uline{RosePO} (CoRR '24)~\citep{liao2024rosepo}, and \uline{SPRec} (WWW '25)~\citep{gao2025sprec}.
    \item \textbf{DPO Methods for LLM-based RS}: direct preference optimization (DPO)~\citep{rafailov2023direct} methods that model pairwise preferences for LLM-based RS, including \uline{S-DPO} (NeurIPS '24)~\citep{chen2024softmax}, \uline{RosePO} (CoRR '24)~\citep{liao2024rosepo}, and \uline{SPRec} (WWW '25)~\citep{gao2025sprec}.
\end{itemize}

\noindentparnoline{Implementation Details.} For conventional sequential recommendation methods, we closely follow their official implementations and default hyperparameters, training a 2-layer and 64-dim transformer for 200 epochs with a batch size of 128 and a learning rate of 1e-3. For LLM-enhanced RS methods, we directly utilize their official code, where SASRec is adopted as the backbone. For LLM-based RS methods, we employ Llama-3.2-3B~\citep{grattafiori2024llama} as the base recommendation model in the main experiments (we also evaluate BEAR's performance on other LLM sizes in \cref{tab:bigrec_llmsize}, including Llama-3.2-1B and Llama-3-8B~\citep{grattafiori2024llama}). Following the recent works~\citep{wang2025msl}, we fine-tune the LLMs for 10 epochs and evaluate the checkpoints with the best NDCG@10 results on the validation set. During training, the standard LoRA technique~\citep{hu2022lora} and AdamW~\citep{loshchilov2017decoupled} optimizer are applied, with a rank of 8 and a learning rate of 1e-4 (except for SPRec, which uses 2e-5). The batch size is set to 64 for DPO methods, and 128 for others. During inference, following previous works~\citep{bao2025bi,bao2024decoding,wang2025msl}, we use constrained beam search with a beam size $B$ of 10, ensuring that the generated items are valid and non-hallucinated. For evaluation, we adopt the widely-used NDCG@$K$ and HitRatio@$K$ metrics to assess the top-$K$ recommendation performance, with $K \in \{5, 10\}$. For our BEAR method, we search hyperparameters $\lambda \in [0.1, 5.0]$ and $\xi \in [0.25, 3.0]$. For all the baseline SFT and DPO-based fine-tuning methods, we closely follow their official implementations and recommended hyperparameters. For hardware, all experiments are conducted on $8 \times$ GeForce RTX 5090 GPUs.

% END: EXPERIMENTS: EXPERIMENTAL SETUP ------------------------------
%%%%%%%%%%%%%%%%%%%%%%%%%%%%%%%%%%%%%%%%%%%%%%%%%%%%%%%%%%%%%%%%%%%%%

% TABLE: DIFFERENT LLM SIZES ----------------------------------------
\begin{table*}[t]
  \centering
  \caption{
    Performance comparison on different LLMs, including Llama-3.2-1B and Llama-3-8B~\citep{grattafiori2024llama}, on the BIGRec~\citep{bao2025bi} backbone. We adopt MSL~\citep{wang2025msl} as the representative SFT method. N@$K$ and H@$K$ denote NDCG@$K$ and HitRatio@$K$ metrics, respectively. The best results are highlighted in bold. "\textcolor{darkred}{\textbf{Imp.\%}}" indicates the relative improvement of BEAR over the best baseline.
  }
  \resizebox{\textwidth}{!}{
    \begin{tabular}{l|cccc|cccc|cccc|cccc}
    \toprule
    \multicolumn{1}{c|}{\multirow{2}[4]{*}{\textbf{Methods}}} 
        & \multicolumn{4}{c|}{\textbf{Office}} 
        & \multicolumn{4}{c|}{\textbf{Book}} 
        & \multicolumn{4}{c|}{\textbf{Toy}} 
        & \multicolumn{4}{c}{\textbf{Clothing}} \\
    \cmidrule(lr){2-5} \cmidrule(lr){6-9} \cmidrule(lr){10-13} \cmidrule(lr){14-17}      
        & \textbf{N@5} & \textbf{N@10} & \textbf{H@5} & \textbf{H@10} 
        & \textbf{N@5} & \textbf{N@10} & \textbf{H@5} & \textbf{H@10} 
        & \textbf{N@5} & \textbf{N@10} & \textbf{H@5} & \textbf{H@10} 
        & \textbf{N@5} & \textbf{N@10} & \textbf{H@5} & \textbf{H@10} \\
    \midrule
    % --- LLM size: 1B ---
    BIGRec (1B) & 0.0095 & 0.0177 & 0.0203 & 0.0447 & 0.0058 & 0.0066 & 0.0114 & 0.0137 & 0.0088 & 0.0133 & 0.0159 & 0.0296 & 0.0032 & 0.0066 & 0.0068 & 0.0176 \\
    + MSL       & 0.0201 & 0.0232 & 0.0322 & 0.0421 & 0.0070 & 0.0089 & 0.0116 & 0.0176 & 0.0165 & 0.0230 & 0.0267 & 0.0465 & 0.0071 & 0.0105 & 0.0136 & 0.0244 \\
    % \midrule
    \textbf{+ BEAR} & \textbf{0.0203} & \textbf{0.0270} & \textbf{0.0343} & \textbf{0.0556} 
        & \textbf{0.0076} & \textbf{0.0096} & \textbf{0.0132} & \textbf{0.0194} 
        & \textbf{0.0176} & \textbf{0.0240} & \textbf{0.0284} & \textbf{0.0484} 
        & \textbf{0.0075} & \textbf{0.0116} & \textbf{0.0144} & \textbf{0.0270} \\
    \textcolor{darkred}{\textbf{Imp.\%}} 
        & \textcolor{darkred}{\textbf{+1.00\%}} & \textcolor{darkred}{\textbf{+16.38\%}} & \textcolor{darkred}{\textbf{+6.52\%}} & \textcolor{darkred}{\textbf{+24.38\%}}
        & \textcolor{darkred}{\textbf{+8.57\%}} & \textcolor{darkred}{\textbf{+7.87\%}} & \textcolor{darkred}{\textbf{+13.79\%}} & \textcolor{darkred}{\textbf{+10.23\%}}
        & \textcolor{darkred}{\textbf{+6.67\%}} & \textcolor{darkred}{\textbf{+4.35\%}} & \textcolor{darkred}{\textbf{+6.37\%}} & \textcolor{darkred}{\textbf{+4.09\%}}
        & \textcolor{darkred}{\textbf{+5.63\%}} & \textcolor{darkred}{\textbf{+10.48\%}} & \textcolor{darkred}{\textbf{+5.88\%}} & \textcolor{darkred}{\textbf{+10.66\%}} \\
    \midrule
    % --- LLM size: 8B ---
    BIGRec (8B) & 0.0114 & 0.0213 & 0.0213 & 0.0514 & 0.0109 & 0.0137 & 0.0169 & 0.0258 & 0.0142 & 0.0189 & 0.0213 & 0.0357 & 0.0057 & 0.0081 & 0.0104 & 0.0180 \\
    + MSL       & 0.0392 & 0.0417 & 0.0556 & 0.0634 & 0.0103 & 0.0135 & 0.0160 & 0.0260 & 0.0245 & 0.0298 & 0.0373 & 0.0538 & 0.0085 & 0.0128 & 0.0156 & 0.0292 \\
    % \midrule
    \textbf{+ BEAR} & \textbf{0.0407} & \textbf{0.0438} & \textbf{0.0608} & \textbf{0.0706} 
        & \textbf{0.0140} & \textbf{0.0172} & \textbf{0.0256} & \textbf{0.0356} 
        & \textbf{0.0262} & \textbf{0.0318} & \textbf{0.0401} & \textbf{0.0572} 
        & \textbf{0.0090} & \textbf{0.0135} & \textbf{0.0162} & \textbf{0.0302} \\
    \textcolor{darkred}{\textbf{Imp.\%}} 
        & \textcolor{darkred}{\textbf{+3.83\%}} & \textcolor{darkred}{\textbf{+5.04\%}} & \textcolor{darkred}{\textbf{+9.35\%}} & \textcolor{darkred}{\textbf{+11.36\%}}
        & \textcolor{darkred}{\textbf{+28.44\%}} & \textcolor{darkred}{\textbf{+25.55\%}} & \textcolor{darkred}{\textbf{+51.48\%}} & \textcolor{darkred}{\textbf{+36.92\%}}
        & \textcolor{darkred}{\textbf{+6.94\%}} & \textcolor{darkred}{\textbf{+6.71\%}} & \textcolor{darkred}{\textbf{+7.51\%}} & \textcolor{darkred}{\textbf{+6.32\%}}
        & \textcolor{darkred}{\textbf{+5.88\%}} & \textcolor{darkred}{\textbf{+5.47\%}} & \textcolor{darkred}{\textbf{+3.85\%}} & \textcolor{darkred}{\textbf{+3.42\%}} \\
    \bottomrule
    \end{tabular}
  }
  \label{tab:bigrec_llmsize}
\end{table*}
% END TABLE: DIFFERENT LLM SIZES ------------------------------------

% FIGURE: S-DPO+BEAR ------------------------------------------------
\begin{figure*}[t]
    \centering
    \includegraphics[width=\textwidth]{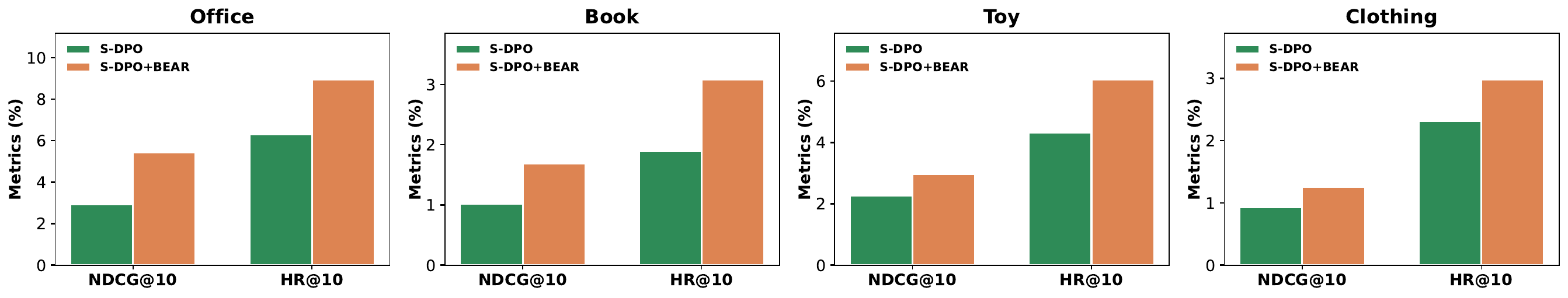}
    \caption{
        Performance of applying BEAR to S-DPO~\citep{chen2024softmax}. In addition to the original S-DPO loss, we apply the positive BEAR regularization to the positive items and the negative BEAR regularization to the sampled negative items. Results show that BEAR significantly enhances the performance of S-DPO across all datasets.
    }
    \Description{Performance Comparison on applying BEAR to S-DPO.}
    \label{fig:sdpo_bear}
\end{figure*}
% END FIGURE: PRUNING RATE ------------------------------------------

%%%%%%%%%%%%%%%%%%%%%%%%%%%%%%%%%%%%%%%%%%%%%%%%%%%%%%%%%%%%%%%%%%%%%
% EXPERIMENTS: EXPERIMENTAL RESULTS ---------------------------------
\subsection{Experimental Results} \label{sec:experiments:results}

In the following, we present and analyze the experimental results to answer the aforementioned research questions.

\noindentparnoline{Overall Performance (RQ1).}
\cref{tab:main-results} presents the overall recommendation performance comparison of BEAR with baselines. As shown, BEAR consistently outperforms all compared methods across all datasets, achieving \emph{an average improvement of +12.50\%} in NDCG@$K$ and HitRatio@$K$ over state-of-the-art SFT methods (\eg MSL) and DPO methods (\eg RosePO). This demonstrates the effectiveness of BEAR in enhancing recommendation quality, which aligns well with our motivation of bridging the training-inference gap in LLM-based RS (\cf \cref{sec:methodology:bear}).

\noindentparnoline{Mitigation of Training-Inference Inconsistency (RQ2).}
To validate BEAR's effectiveness in reducing incorrect pruning of positive items during beam search, we compare pruning rate metrics (\ie PR@$K$ in \cref{fig:pruning-rate}) between BEAR and SFT methods (\eg MSL). Specifically, PR@$K$ measures the proportion of positive items that are ranked within the top-$K$ based on their overall probabilities but are incorrectly pruned during beam search. As shown in \cref{fig:pruning-rate}, BEAR exhibits a significantly lower PR@$K$, achieving \emph{an average reduction of -24.86\%} compared to SFT methods. This verifies that our beam-search-aware regularization \eqnref{eq:beam_aware_regularization} can effectively reduce the risk of incorrectly pruning positive items, thereby mitigating the training-inference inconsistency in LLM-based RS.

\noindentparnoline{Generalizability: Backbones and LLM Sizes (RQ3-1).}
To assess the generalizability of BEAR, we further evaluate its performance on different LLM-based RS backbones (\eg LLaRA~\citep{liao2024llara} and A-LLMRec~\citep{kim2024large}) with different base LLM sizes (\eg Llama-3.2-1B and Llama-3-8B~\citep{grattafiori2024llama}). As illustrated in \cref{tab:different-backbones} and \cref{tab:bigrec_llmsize}, BEAR consistently outperforms the SFT methods across different backbones and LLM sizes, demonstrating its versatility and effectiveness.

\noindentparnoline{Generalizability: Fine-tuning Objectives (RQ3-2).}
Given that the training-inference inconsistency is a common challenge in fine-tuning LLMs for recommendation, exploring the applicability of BEAR to other fine-tuning objectives, \eg DPO-based methods, is also of great interest. Therefore, we also apply BEAR to the widely-adopted S-DPO~\citep{chen2024softmax}. Specifically, in addition to the original S-DPO loss, we apply the \emph{positive} BEAR regularization in \eqnref{eq:beam_aware_regularization} (\ie $\mathcal{L}_{\textnormal{reg}}$) to the positive items, while applying the \emph{negative} regularization (\ie $-\mathcal{L}_{\textnormal{reg}}$) to the sampled negative items. This encourages the model to retain positive items during beam search and prune negative items, thus aligning the DPO training with the inference process. As shown in \cref{fig:sdpo_bear}, BEAR significantly enhances the performance of S-DPO, validating its broad applicability in improving various fine-tuning objectives for LLM-based RS.

\begin{figure}[t]
    \centering
    \includegraphics[width=\columnwidth]{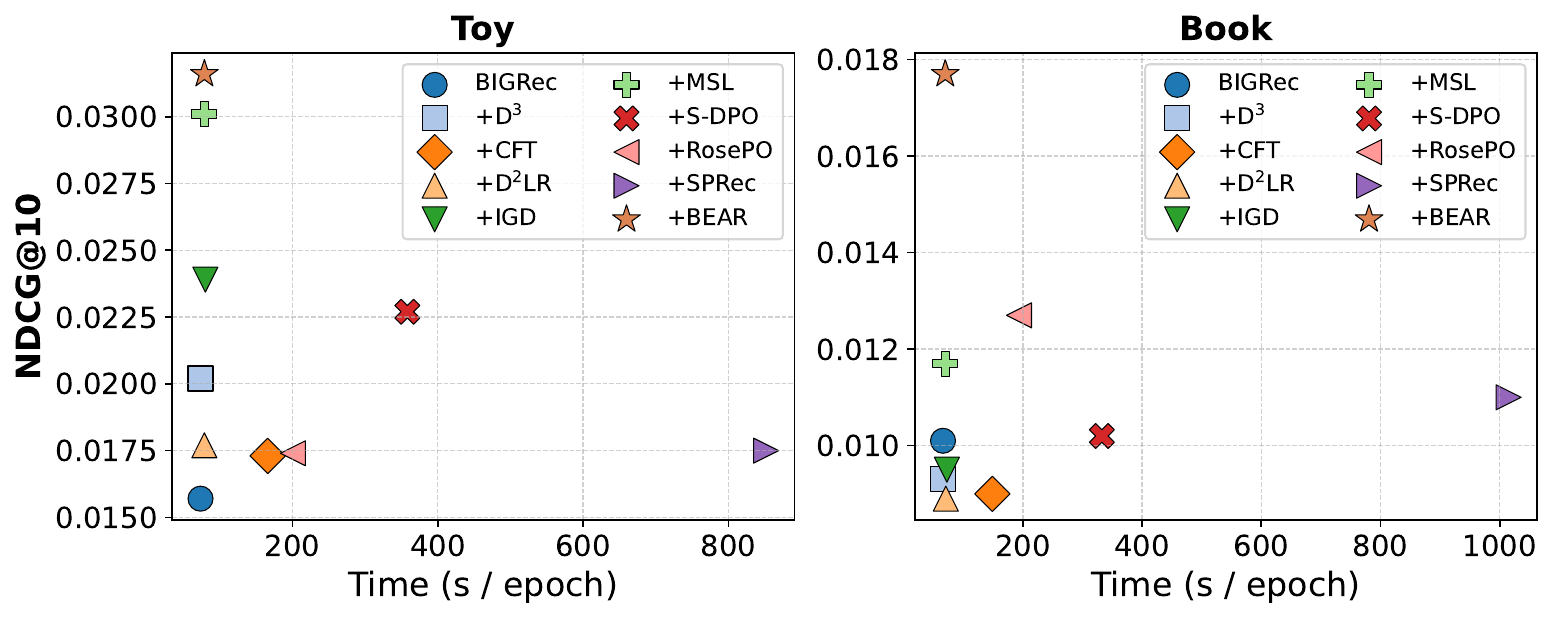}
    \caption{
        Training efficiency comparison between BEAR and other baselines. The x-axis represents the training time per epoch, and the y-axis represents the NDCG@10 performance.
    }
    \Description{Training Efficiency Comparison between BEAR and Other LLM-based RS Methods.}
    \label{fig:training-efficiency}
\end{figure}

\noindentparnoline{Computational Efficiency (RQ4).}
As illustrated in \cref{fig:training-efficiency}, BEAR demonstrates both comparable training efficiency to SFT methods and optimal recommendation performance among all compared methods. This is attributable to BEAR's meticulous design, which avoids additional forward passes of the LLM during training, as discussed in \cref{sec:methodology:bear}. In contrast, other methods such as CFT~\citep{zhang2024causality}, S-DPO~\citep{chen2024softmax}, and SPRec~\citep{gao2025sprec} require additional forward passes for each training instance, leading to noteworthy increased computational overhead (typically $2 \times$ to $10 \times$ compared to BEAR). This inefficiency hinders their scalability in real-world applications, making BEAR a more practical choice for deploying LLM-based recommender systems.

\begin{figure}[t]
    \centering
    \includegraphics[width=\columnwidth]{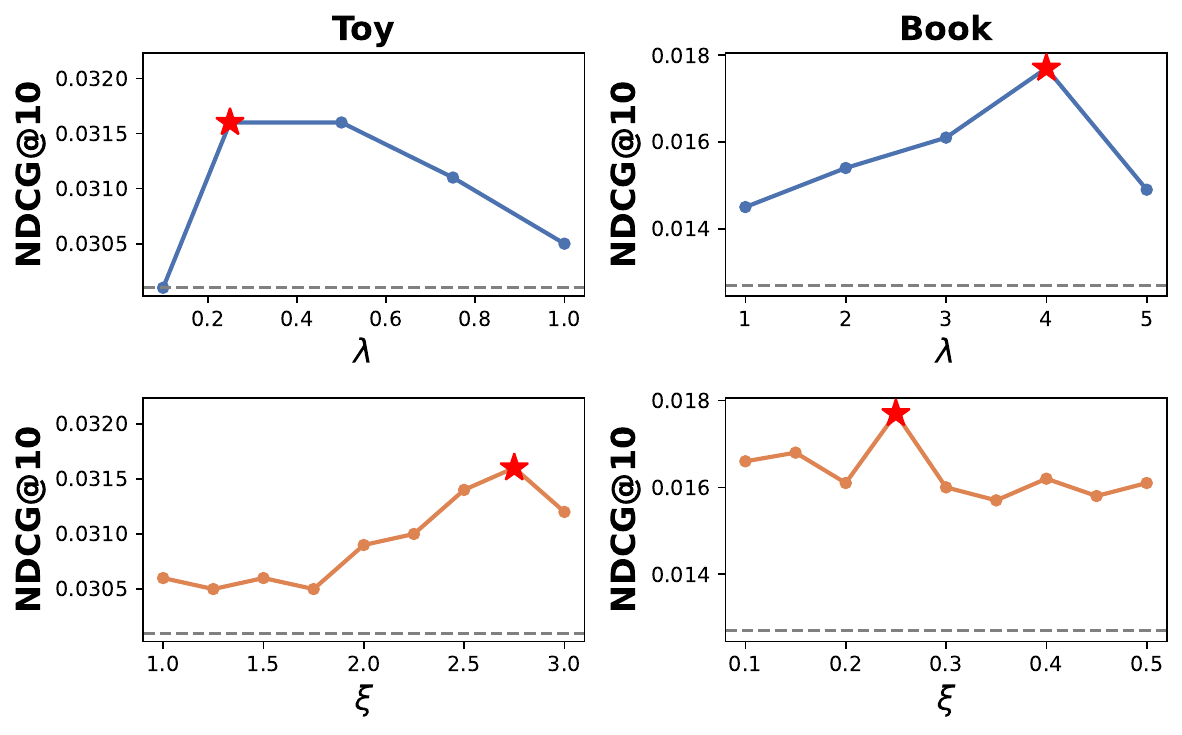}
    \caption{
        Sensitivity analysis of BEAR's hyperparameters $\lambda$ (\textcolor{sensitivity_blue}{blue}) and $\xi$ (\textcolor{sensitivity_orange}{orange}). The best result is highlighted with a star, and the baseline result is indicated with a dashed line. 
    }
    \Description{Hyperparameter Sensitivity Analysis of BEAR.}
    \label{fig:sensitivity}
\end{figure}

\noindentparnoline{Hyperparameter Sensitivity (RQ5).}
In \cref{fig:sensitivity}, we conduct a sensitivity analysis of BEAR's hyperparameters $\lambda$ and $\xi$. As illustrated, the performance for both $\lambda$ and $\xi$ exhibits a clear peak. Specifically, the regularization weight $\lambda$ serves to balance the SFT and the regularization: when $\lambda$ is too small, BEAR degenerates to standard SFT, leading to suboptimal performance due to the training-inference inconsistency; conversely, an excessively large $\lambda$ overemphasizes the regularization, potentially failing to maximize the overall probability of positive items. For the temperature parameter $\xi$, it controls the smoothness of the approximation to the indicator function: when $\xi$ is too small, the approximation becomes overly sharp, resulting in gradient saturation; on the other hand, a too-large $\xi$ leads to an overly smooth approximation, diluting the regularization signal. In addition, we observe that BEAR is relatively robust to hyperparameter variations within a reasonable range, which is beneficial for practical applications.

% END: EXPERIMENTS: EXPERIMENTAL RESULTS ----------------------------
%%%%%%%%%%%%%%%%%%%%%%%%%%%%%%%%%%%%%%%%%%%%%%%%%%%%%%%%%%%%%%%%%%%%%

% END: EXPERIMENTS --------------------------------------------------
%%%%%%%%%%%%%%%%%%%%%%%%%%%%%%%%%%%%%%%%%%%%%%%%%%%%%%%%%%%%%%%%%%%%%

% FIGURE: HYPER-PARAMETER SENSITIVITY -------------------------------
% END FIGURE: HYPER-PARAMETER SENSITIVITY ---------------------------

%%%%%%%%%%%%%%%%%%%%%%%%%%%%%%%%%%%%%%%%%%%%%%%%%%%%%%%%%%%%%%%%%%%%%
% RELATED WORK ------------------------------------------------------
\section{Related Work} \label{sec:related_work}

\noindentparnolinenospace{Sequential Recommendation.}
Sequential recommendation~\citep{chang2021sequential,chen2018sequential,li2020time,wang2024llm4dsr} has long been a central task in recommendation research, aiming to predict a user's next interaction based on historical behavior sequences. Classic models such as GRU4Rec~\citep{hidasi2015session}, Caser~\citep{tang2018personalized}, SASRec~\citep{kang2018self}, and BERT4Rec~\citep{sun2019bert4rec} have achieved remarkable success by leveraging different architectures to model temporal dependencies~\citep{hochreiter1997long,krizhevsky2012imagenet,vaswani2017attention}. In recent years, by scaling up model sizes to billions of parameters~\citep{zhai2024actions}, integrating multi-side information~\citep{xia2023transact,zhou2023bootstrap,wei2023multi}, performing contextual reasoning~\citep{tang2025think,liu2025lares}, and enabling continual learning~\citep{mi2020ader,cai2022reloop}, modern sequential recommendation models have significantly advanced the capabilities of recommender systems.

\noindentparnoline{LLM-based Recommender Systems.}
Large Language Models (LLMs)~\citep{achiam2023gpt,bai2023qwen,touvron2023llama} have shown exceptional generative~\citep{brown2020language}, reasoning~\citep{wei2022chain}, and generalization~\citep{radford2019language} abilities, inspiring research on their integration into recommender systems (\ie LLM-based RS)~\citep{hou2024large,liu2024once,wang2023zero,gao2023chat,liu2023chatgpt,dai2023uncovering}. In this paradigm, recommendation is typically reformulated as a natural language task, where the LLMs are adapted to generate the textual descriptions of the recommended items based on the user history and instructions, thereby leveraging their powerful capabilities for recommendation~\citep{wang2024recommend,bao2023tallrec,zhang2025recommendation,li2023exploring,shi2024large,geng2022recommendation,cui2022m6,kim2024large,liao2024llara}.

Recently, a series of optimization methods have been proposed to enhance the performance of LLM-based RS. Following the standard practice in NLP, supervised fine-tuning (SFT)~\citep{ouyang2022training,bao2025bi,bao2024decoding} is widely adopted, aiming to maximize the positive item likelihood. Various enhancements to the basic SFT have been proposed, focusing on causality~\citep{zhang2024causality}, popularity~\citep{lu2025dual}, entropy~\citep{lin2025igd}, and masking~\citep{wang2025msl}. Built upon SFT, some researchers have also explored variants of direct preference optimization (DPO)~\citep{rafailov2023direct}, which contrasts preferred and rejected items to model pairwise preferences, including strategies like multiple negatives ranking~\citep{chen2024softmax}, negative sampling strategies~\citep{liao2024rosepo}, and self-play mechanism~\citep{gao2025sprec}.

Despite these promising results, existing fine-tuning strategies still face significant limitations due to their ignorance of the training-inference inconsistency (\cf \cref{sec:preliminaries:limitations-of-sft}). Specifically, they typically optimize the overall probability of positive items, which does not guarantee that these items will be retained during the greedy pruning of beam search. This discrepancy can lead to suboptimal recommendation performance, as the final beam search candidates may not include the most relevant items. Targeting this issue, our proposed BEAR method tries to bridge this inconsistency by explicitly considering the beam search process during training, thereby enhancing the recommendation performance of LLM-based RS.

\noindentparnoline{Related Work on Generative Retrieval.}
Recent studies in other domains, such as generative retrieval~\citep{li2025matching,wu2025constrained}, have explored similar training-inference inconsistency issues caused by beam search. These works can be categorized into three main approaches. The first approach, \eg OTM~\citep{zhuo2020learning} and DTR~\citep{liu2024learning}, focuses on learning a tree model that is Bayes optimal for beam search, or equivalently, satisfies the max-heap property (\ie each parent node's preference should be the maximum among its children). The Bayes optimality ensures that the items with top-$B$ overall probabilities will never be pruned. However, transferring this idea to LLM-based RS is non-trivial, as training such a tree model typically requires performing beam search during training, which is computationally impractical for LLMs, as discussed in \cref{sec:methodology:sufficient}. The second approach, \eg RIPOR~\citep{zeng2024scalable}, involves optimizing the prefix probabilities of positive items against negatives to reduce incorrect pruning, which is similar to the objective \eqnref{eq:sufficient_condition}. Unfortunately, these methods rely on a large amount of negative sampling, incurring over a 100$\times$ computational overhead, rendering it impractical for LLM-based RS. The third approach, \eg ProphetNet-Ads~\citep{qi2020prophetnet} and PAG~\citep{zeng2024planning}, heuristically incorporates future token or prior item scores to guide beam search. Nonetheless, accurately estimating these look-ahead scores is equivalent to learning a precise recommender, which is inherently difficult. Different from these strategies, our BEAR tackles this issue from a more principled perspective by optimizing a necessary condition for retrieving positive items during beam search, achieving better performance and computational efficiency.

% For instance, RIPOR~\citep{zeng2024scalable} proposes prefix-oriented ranking optimization, which contrasts prefixes of positive items against sampled negatives with multiple prefix lengths, aiming to improve the prefix probabilities and reduce incorrect pruning. However, RIPOR relies on a large amount of negative sampling, incurring over a 100$\times$ computational overhead, rendering it impractical for LLM-based RS. Moreover, PAG~\citep{zeng2024planning} builds upon RIPOR by incorporating the item-level prior scores calculated from set-based DocIDs to guide beam scoring process. Nonetheless, applying PAG to LLM-based RS remains challenging, as accurately estimating these item-level scores is equivalent to learning a precise recommender, which is inherently difficult. Besides, these methods are mostly heuristic and lack theoretical guarantees. Different from these strategies, our BEAR tackles this issue from a more principled perspective by optimizing a necessary condition for retaining positive items during beam search, achieving both theoretical soundness and computational efficiency.

% END: RELATED WORK -------------------------------------------------
%%%%%%%%%%%%%%%%%%%%%%%%%%%%%%%%%%%%%%%%%%%%%%%%%%%%%%%%%%%%%%%%%%%%%

% FIGURE: PRUNING RATE ----------------------------------------------
% END FIGURE: PRUNING RATE ------------------------------------------

% TABLE: SUFFICIENT-VS-NECESSARY ------------------------------------
\begin{table}[t]
  \centering
  \caption{
    Performance comparison between SFT, BEAR, and the straightforward solution that directly optimizes the objective \eqnref{eq:sufficient_condition}, on the BIGRec~\citep{bao2025bi} backbone. Due to the high computational cost of beam search required for optimizing \eqnref{eq:sufficient_condition}, experiments are conducted on two small-scale datasets (Toy-S and Book-S) with 2K sampled users.
  }
  \resizebox{\columnwidth}{!}{
    \begin{tabular}{l|cc|cc}
    \toprule
    \multicolumn{1}{c|}{\multirow{2}[4]{*}{\textbf{Methods}}} 
        & \multicolumn{2}{c|}{\textbf{Toy-S}} 
        & \multicolumn{2}{c}{\textbf{Book-S}} \\
    \cmidrule(lr){2-3} \cmidrule(lr){4-5}      
        & \textbf{NDCG@10} & \textbf{HR@10} 
        & \textbf{NDCG@10} & \textbf{HR@10} \\
    \midrule
    + SFT           & 0.0076 & 0.0200 & 0.0046 & 0.0120 \\
    + BEAR          & 0.0094 & 0.0260 & 0.0072 & 0.0180 \\
    + Obj.~\eqnref{eq:sufficient_condition}    & 0.0103 & 0.0280 & 0.0077 & 0.0200 \\
    \bottomrule
    \end{tabular}
    }
    \label{tab:sufficient-vs-necessary}
\end{table}
% END TABLE: SUFFICIENT-VS-NECESSARY --------------------------------

%%%%%%%%%%%%%%%%%%%%%%%%%%%%%%%%%%%%%%%%%%%%%%%%%%%%%%%%%%%%%%%%%%%%%
% CONCLUSION AND FUTURE DIRECTIONS ----------------------------------
\section{Conclusion and Future Directions} \label{sec:conclusion_and_future_directions}

In this paper, we identify and address the training-inference inconsistency in LLM-based recommender systems, which arises from the discrepancy between the SFT training objective and the greedy pruning nature of beam search. To bridge this gap, we propose BEAR, a novel beam-search-aware regularization method that explicitly considers the beam search behavior during training. BEAR optimizes a necessary condition for retaining positive items during beam search, thereby reducing the risk of incorrect pruning and enhancing recommendation performance, while incurring minimal computational overhead compared to standard SFT. For future directions, exploring the integration of BEAR with inference-stage techniques, such as multi-token prediction~\citep{gloeckle2024better} and diverse beam search~\citep{vijayakumar2018diverse}, would be a promising avenue.

% END: CONCLUSION AND FUTURE DIRECTIONS -----------------------------
%%%%%%%%%%%%%%%%%%%%%%%%%%%%%%%%%%%%%%%%%%%%%%%%%%%%%%%%%%%%%%%%%%%%%

%%%%%%%%%%%%%%%%%%%%%%%%%%%%%%%%%%%%%%%%%%%%%%%%%%%%%%%%%%%%%%%%%%%%%
% ACKNOWLEDGMENTS ---------------------------------------------------

% The acknowledgments section is defined using the "acks" environment
% (and NOT an unnumbered section). This ensures the proper
% identification of the section in the article metadata, and the
% consistent spelling of the heading.
\begin{acks}
    This work is supported by the Starry Night Science Fund of Zhejiang University Shanghai Institute for Advanced Study (SN-ZJU-SIAS-001) and the National Natural Science Foundation of China (62476244, 62372399). This work is also funded by ZJU-China Unicom Digital Security Joint Laboratory.
\end{acks}

% END: ACKNOWLEDGMENTS ----------------------------------------------
%%%%%%%%%%%%%%%%%%%%%%%%%%%%%%%%%%%%%%%%%%%%%%%%%%%%%%%%%%%%%%%%%%%%%

%%%%%%%%%%%%%%%%%%%%%%%%%%%%%%%%%%%%%%%%%%%%%%%%%%%%%%%%%%%%%%%%%%%%%
% APPENDICES --------------------------------------------------------
% \clearpage
\appendix

\section{Appendix: Further Discussions} \label{app:further_discussions}

\noindentparnolinenospace{BEAR \versus Straightforward Solution.}
As discussed in \cref{sec:methodology:sufficient}, a naive solution to allievate the incorrect pruning issue is directly optimizing the objective \eqnref{eq:sufficient_condition}. While this straightforward approach seems "theoretically optimal", it requires performing beam search during training, incurring impractical computational overhead in LLM-based RS. Instead, our proposed BEAR method opts to optimize a necessary condition for retaining positive items, offering a more computationally efficient alternative. A crucial research question naturally arises: \emph{Does optimizing this weaker, necessary condition sacrifice significant effectiveness compared to the "optimal" yet expensive solution?}

To answer this, we conduct experiments comparing BEAR with directly optimizing \eqnref{eq:sufficient_condition} on two datasets in \cref{tab:sufficient-vs-necessary}. As observed, both BEAR and the computationally intensive objective \eqnref{eq:sufficient_condition} significantly outperform standard SFT, validating their ability to mitigate training-inference inconsistency. More importantly, the performance gap between BEAR and objective \eqnref{eq:sufficient_condition} is marginal. This indicates that BEAR is \emph{sufficiently effective} at capturing the core dynamics of beam search pruning, achieving performance parity with the straightforward solution while avoiding its severe computational costs. Notably, these results also align perfectly with our finding in \cref{fig:pr_and_tpr_pruning_rate}: since violating the necessary condition accounts for the vast majority of incorrect pruning cases, effectively targeting this primary cause via BEAR yields the majority of the potential gains, which further explains its strong performance.

\bibliographystyle{ACM-Reference-Format}
% NOTE: add \balance to end of the references to balance the columns
\balance
\bibliography{references}

% END: REFERENCES ---------------------------------------------------
%%%%%%%%%%%%%%%%%%%%%%%%%%%%%%%%%%%%%%%%%%%%%%%%%%%%%%%%%%%%%%%%%%%%%

\end{document}